\documentclass[10pt,journal]{IEEEtran}

\usepackage{blkarray}%
\usepackage{graphicx}%
\usepackage{amsmath}
\usepackage{amssymb}
\usepackage{amsfonts}
\usepackage{amsthm}
\usepackage[mathcal]{eucal}
\usepackage{mathrsfs}
\usepackage{booktabs}
\usepackage{enumerate}
\usepackage{multirow}
\usepackage[subrefformat=parens,farskip=0pt,justification=centering]{subfig}
\usepackage{color}
\usepackage{cite}%
\usepackage{comment}%
\usepackage{soul}%
\soulregister\cite7
\soulregister\ref7
\soulregister\pageref7
\usepackage{etoolbox}%
\usepackage{url}
\usepackage{nth}%
\usepackage{bm}%
\usepackage{courier}
\usepackage{balance}
\usepackage{threeparttable}
\usepackage{xcolor,colortbl}
\usepackage{footnote}
\usepackage{listings}
\usepackage{setspace}%
\usepackage[inline]{enumitem}
\usepackage{verbatim}
\usepackage[bookmarks=false]{hyperref}
\hypersetup{
    colorlinks = true,
    citecolor  = blue,
    linkcolor  = blue,
    urlcolor   = blue,
}
\usepackage{tikz}
\usetikzlibrary{patterns,snakes}
\usetikzlibrary{positioning,calc,fit,decorations.pathmorphing,shapes.geometric, shapes.gates.logic.US, calc}
\usetikzlibrary{arrows,arrows.meta,decorations.markings,shapes,shapes.arrows}
\usetikzlibrary{decorations,decorations.pathreplacing}
\usetikzlibrary{backgrounds}
\usepackage{filecontents}%
\usepackage{pgfplots}
\usepackage{pgfplotstable}
\usepackage{scalefnt}
\pgfplotsset{compat=newest}
\usepackage{caption}
\usepackage{pifont}%
\usepackage{cleveref}
\Crefformat{figure}{Fig.~#2#1#3}%
\Crefname{subfigure}{Fig.}{Figs.}
\Crefname{figure}{Fig.}{Figs.}
\Crefformat{table}{TABLE~#2#1#3}%
\usepackage[figuresright]{rotating}

\usepackage{algorithm}
\iftrue
\usepackage{algpseudocode}%
\algrenewcommand\textproc{\texttt}
\makeatletter
\let\OldStatex\Statex
\renewcommand{\Statex}[1][3]{%
  \setlength\@tempdima{\algorithmicindent}%
  \OldStatex\hskip\dimexpr#1\@tempdima\relax
}
\makeatother
\else
\usepackage{algorithmic}
\fi

\definecolor{CUHKorange}{RGB}{244,106,18}%
\definecolor{CUHKblue}{RGB}{0,111,190}%
\definecolor{CUHKgreen}{RGB}{0,127,128}%
\definecolor{CUHKred}{RGB}{228,46,36}%
\definecolor{CUHKyellow}{RGB}{198,148,34}%
\definecolor{CUHKdark}{RGB}{114,44,114}%
\definecolor{CUHKmiddle}{RGB}{144,44,144}%
\definecolor{CUHKlight}{RGB}{167,44,167} 
\definecolor{CUHKpurple}{RGB}{117,15,109}
\definecolor{CUHKgold}{RGB}{221,163,0}
\definecolor{CUHKribbon}{RGB}{244,223,176}
\definecolor{CUHKblack}{RGB}{34,24,21}

\renewcommand{\bf}[1]{\textbf{#1}}
\renewcommand{\tt}[1]{\texttt{#1}}

\usepackage{tcolorbox}
\tcbuselibrary{skins,breakable}
    {\endtcolorbox}
    {\endtcolorbox}

\crefname{mytheorem}{Theorem}{Theorems}
\crefname{mylemma}{Lemma}{Lemmas}
\crefname{myclaim}{Claim}{Claims}
\crefname{myproperty}{Property}{Properties}
\crefname{mycorollary}{Corollary}{Corollaries}

\RequirePackage[normalem]{ulem}%
\RequirePackage{color}\definecolor{RED}{rgb}{1,0,0}\definecolor{BLUE}{rgb}{0,0,1}%

\graphicspath{{./figs/}{../}}

\usepackage[round-pad=false, round-mode=places,round-precision=2,group-separator={,},output-decimal-marker={.}]{siunitx}
\usepackage{mathtools}

\begin{document}

\title{
    Rect3D: A Unified Analytical Framework\\for 3D-IC Rectilinear Floorplanning
}

\author{Shuo~Ren, Rongliang~Fu${^*}$, Libo~Shen, Zhen~Zhuang, Leilei~Jin,\\
Chen~Wu, Lei~He~\IEEEmembership{Fellow,~IEEE}, Bei~Yu~\IEEEmembership{Senior Member,~IEEE}, Tsung-Yi~Ho~\IEEEmembership{Fellow,~IEEE}
    \thanks{This work was conducted in the JC STEM Lab of Intelligent Design Automation, funded by The Hong Kong Jockey Club Charities Trust. This work was supported in part by the Research Grants Council of the Hong Kong Special Administrative Region, China, under Grant Nos. CUHK14211324 and CUHK7010840, and in part by ACCESS - AI Chip Center for Emerging Smart Systems, supported by the InnoHK initiative of the Innovation and Technology Commission of the Hong Kong Special Administrative Region Government.}
    \IEEEcompsocitemizethanks{
        \IEEEcompsocthanksitem Shuo~Ren, Rongliang~Fu, Libo~Shen, Zhen Zhuang, Leilei~Jin, Bei Yu, and Tsung-Yi~Ho are with The Chinese University of Hong Kong.
        E-mail: \{sren, rlfu, lbshen24, zzhuang21, lljin, ~byu,~tyho\}@cse.cuhk.edu.hk.
        \IEEEcompsocthanksitem Chen~Wu and Lei~He are with Eastern Institute of Technology, Ningbo.
        E-mail: cwu@idt.eitech.edu.cn and lhe@eitech.edu.cn.
        \IEEEcompsocthanksitem  ${^*}$Corresponding author: Rongliang Fu.
    }
}

\markboth{Ren \MakeLowercase{\textit{et al.}}: Rect3D: A Unified Analytical Framework for 3D-IC Rectilinear Floorplanning}%
{Ren \MakeLowercase{\textit{et al.}}: Rect3D}

\maketitle
\thispagestyle{plain}
\pagestyle{plain}

\begin{abstract}
3D-ICs offer significant performance improvements for modern VLSI designs by reducing global interconnect cost.
However, conventional 3D floorplanning methods decompose the problem into separate inter-die partitioning and intra-die floorplanning stages, which can restrict the design optimization space and limit the potential gains.
Although directly modeling and optimizing in 3D space can mitigate this limitation, the high computational complexity hinders algorithmic efficiency.
To address these challenges, we propose \textsc{Rect3D}, an analytical 3D rectilinear floorplanning framework that integrates probabilistic inter-die block assignment into a unified continuous optimization model. The framework combines graph Laplacian initialization for topology-aware seeding, a scalable gradient-based global optimization procedure for joint die assignment and geometric refinement, and a 3D grid-based legalization method for generating connected rectilinear layouts.
On GSRC benchmarks, \textsc{Rect3D} reduces wirelength by up to 83.6\% and runtime by up to 15.98$\times$ compared with representative state-of-the-art 3D floorplanning baselines. It also consistently achieves the lowest wirelength among six additional partition-first rectilinear baselines, showing the advantage of preserving die assignment and in-die geometry in a unified 3D optimization flow.
\end{abstract}

\begin{IEEEkeywords}
3D integration, floorplanning, rectilinear, graph Laplacian, analytical optimization.
\end{IEEEkeywords}

\section{Introduction}
\label{sec:intro}

While device scaling has continued to sustain Moore's Law, interconnect scaling has lagged behind, and routing delay has become the dominant bottleneck in modern 2D-ICs~\cite{2019JSA-Cao}. A practical remedy is 3D integration, which stacks multiple device layers so that long intra-die wires are replaced by short vertical connections and overall integration density increases~\cite{2025ASPDAC-towards3dic}. Recent advances in fine-pitch hybrid bonding have further accelerated the deployment of 3D integration in commercial systems~\cite{2024ECTC-hybrid-bonding,2024-cucu-hybridbonding,hybrid-cores-intel}. This momentum is also reflected in ongoing progress in bonding infrastructure and broader 3D integration roadmaps~\cite{2023ECTC-Netzband,2023ECTC-hybrid-bonding-process,ISPD24-Liu,zhou2024research}. As a result, dedicated 3D floorplanning has become increasingly important for fully exploiting the benefits of vertical integration.

However, existing 3D floorplanning methods still leave an important gap
between flexibility and scalability. Broadly speaking, these methods can be
divided into two families.
One family does not fix the inter-die partition in
advance, but instead allows blocks to keep adjusting their two-die assignment
while their in-die geometry is optimized jointly throughout the search. Representative examples
include 3D slicing trees, partition-and-sequence pairs, grouped sequence-pair
formulations, direct 3D extensions, and more recent
learning-based methods~\cite{2005ASPDAC-flpdatastructure-3dslicingtree,2010Integration-Partition-Sequence-pair,2011TVLSI-grouped-sequence-pair,2021TVLSI-thermal-aware-3dfloorplan-tsv,2024neurips-flexplanner}.
The advantage of these methods is that they do not prematurely restrict each
block to the top or bottom die. Instead, a block may move within the die plane
while also choosing between the two dies during optimization. However, this
same flexibility greatly enlarges the search space: each block must be
optimized in both planar geometry and inter-die assignment, which makes the
problem harder to converge and increasingly difficult to solve with high
quality as the design size grows.

Another family pursues better scalability by adopting a partition-first flow.
Representative pipelines first generate a top/bottom split with
Fiduccia-Mattheyses (FM) or spectral
partitioning~\cite{1995ICCAD-FM_partition,1995DAC-Spectral_partitioning,2022TCAD-snap3d}. The resulting per-die subproblems are then
handled by strong 2D rectangular floorplanners based on
Attractor-Repeller-style methods, matrix-based optimization, or
other analytical floorplanning flows~\cite{2008ASPDAC-Luo-AR,2010ASPDAC-Lin-UFO,2023DAC-Li,2026ASPDAC_Ren,2023ICCAD_Huang_orientation_citePef}.
This strategy reuses mature 2D optimization engines and therefore scales
better. However, even with a strong downstream 2D floorplanner, the final
quality is still upper-bounded by the partition result: once the top/bottom
assignment is fixed, later floorplanning can only refine geometry within each
die and has limited ability to recover from poor cross-die decisions.
\begin{figure}[t!]
    \centering
    \includegraphics[width=.85\linewidth]{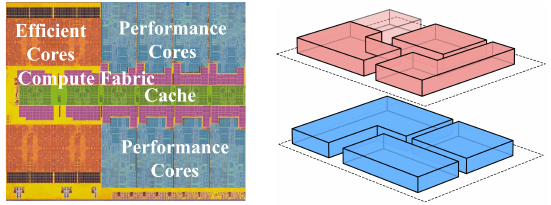}
    \caption{Motivation for 3D rectilinear floorplanning. (Left) Die shot of Intel's Core i9-13900K~\cite{i9_13900K_dieshot_ref} shows commercial adoption of rectilinear modules in modern processors. (Right) Illustration of the proposed 3D rectilinear IC architecture that extends rectilinear floorplanning to vertical integration.}
    \label{fig:3dflp_illustration}
\end{figure}
Moreover, as shown in the Intel Core i9-13900K die shot in~\Cref{fig:3dflp_illustration},
rectilinear module footprints are common in commercial processors. This
supports rectilinear modeling as a practical design abstraction rather than an
idealized one. Recent 2D rectilinear floorplanners such as Modern,
JigsawPlanner, and the ISEDA methods~\cite{2024ICCAD_Chen_Rectilinear_soft,2024ICCAD_Jisaw,2025ISEDA_Rectilinear}
further show that polygonal modules can improve area utilization, but these
methods remain limited to single-die layouts. Extending these strong 2D
rectilinear floorplanners to 3D is therefore not simply a matter of replacing a
rectangular per-die engine with a rectilinear one inside a partition-first
pipeline. The upstream top/bottom split is still committed before the true
cross-die geometric tradeoff becomes visible, so later rectilinear floorplanning
often struggles to correct a poor partition. As our later comparisons show, such
partition-first rectilinear extensions still leave a large quality gap.
\begin{figure}[t!]
    \centering
    \includegraphics[width=.85\linewidth]{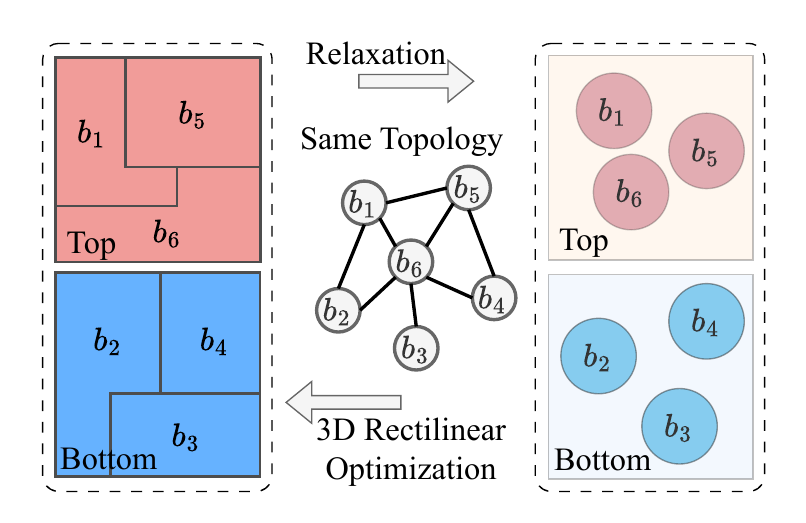}
    \caption{Illustration of topology maintenance: rectilinear blocks are relaxed into circular disks while preserving adjacency relations, enabling spectral embedding in 3D continuous space.}
    \label{fig:topology}
\end{figure}

Taken together, existing approaches either search directly in a large 3D
solution space and become increasingly expensive, or improve scalability by
fixing die assignment too early and thereby restricting the achievable quality.
This tension makes 3D-IC rectilinear floorplanning an important yet
underexplored problem. A more effective solution must optimize die assignment
and in-die geometry jointly, while also preserving the topology-aware
cross-die structure illustrated in~\Cref{fig:topology}. This need motivates a
unified 3D rectilinear floorplanning framework.
The main contributions are summarized as follows:

\begin{itemize}[leftmargin=*, itemsep=2pt]
\item We present \textsc{Rect3D}, to the best of our knowledge the first 3D-IC rectilinear floorplanning framework, which extends beyond conventional rectangular representations to better utilize silicon area and enable native cross-die coordination.

\item We develop a 3D graph Laplacian--based initialization that leverages netlist connectivity, yielding on average 20.7\% shorter HPWL compared to a representative grid-based initialization baseline.

\item We formulate die assignment as a continuous optimization problem where $z$-coordinates encode layer probabilities, and solve the resulting unified 3D objective with an adaptive three-phase regularization strategy and box-constrained L-BFGS-B~\cite{lbfgsb_proposed}, achieving up to 93\% runtime reduction over representative prior 3D methods while simultaneously improving wirelength.

\item \textsc{Rect3D} further introduces a 3D rectilinear legalization method that accounts for rectilinear block constraints across dies, ensuring overlap-free layouts.
\end{itemize}

On GSRC benchmarks, \textsc{Rect3D} achieves up to 83.6\% wirelength reduction and up to 15.98$\times$ speedup over representative state-of-the-art 3D floorplanning baselines. It also consistently achieves the best wirelength among six additional partition-first rectilinear baselines.

The remainder of this paper is organized as follows. \Cref{sec:prelim} defines the problem setting and notation. \Cref{sec:methodology} explains the three stages of \textsc{Rect3D}. \Cref{sec:experiments} shows why each stage matters and how the full framework compares with prior methods. \Cref{sec:conclusion} closes the paper.

\section{Preliminaries}
\label{sec:prelim}

\subsection{Terminologies}
\label{subsec_terminology}

We consider fixed-outline rectilinear floorplanning for two-die face-to-face (F2F) bonded 3D-ICs. In this setting, each soft block is assigned to either the top or bottom die and is ultimately realized as a rectilinear region. The key notation is summarized below.

\begin{itemize}
  \item $G=(\mathcal{B},\mathcal{N})$: the input graph, where $\mathcal{B}=\{b_1, b_2, \dots, b_n\}$ is the set of soft blocks with area $a_i$ for each $b_i$, and $\mathcal{N}$ is the netlist with each net $e \in \mathcal{N}$ assigned a criticality weight $c_e$.
  \item $S_e \subseteq \mathcal{B}$: the set of blocks connected to net $e \in \mathcal{N}$.
  \item $\mathcal{T}$: the set of dies available for block assignment; in the final floorplan each block $b_i$ is assigned to a die in $\mathcal{T}$.
  \item $A \in \mathbb{R}^{n \times n}$: the connection matrix derived from $\mathcal{N}$, 
where each entry is defined as 
$A_{ij} = \sum_{e \in \mathcal{N}:\; \{b_i,b_j\} \subseteq S_e} c_e$, 
following the standard construction in~\cite{2008ASPDAC-Luo-AR,2023DAC-Li}.
  \item $\mathbf{L} = \boldsymbol{\Delta} - A$: the graph Laplacian matrix, where $\boldsymbol{\Delta}$ is the degree matrix with $\Delta_{ii} = \sum_{j} A_{ij}$.
  \item $\mathbf{X}=[\mathbf{x}, \mathbf{y}, \mathbf{z}] \in \mathbb{R}^{n \times 3}$: the coordinate matrix of all blocks, where the three columns correspond to the coordinate vectors $\mathbf{x} = (x_1,\dots,x_n)^\top$, $\mathbf{y} = (y_1,\dots,y_n)^\top$, and $\mathbf{z} = (z_1,\dots,z_n)^\top$. For each block $b_i$, its center coordinate $(x_i,y_i,z_i)$ is given by the $i$-th row of $\mathbf{X}$. During continuous optimization, $z_i \in [0,1]$ acts as a soft die indicator.
\end{itemize}
\subsection{3D IC Floorplanning}
\label{subsec:prelim_3dfloorplan}
\begin{figure*}[t!]
  \centering
  \includegraphics[width=\linewidth]{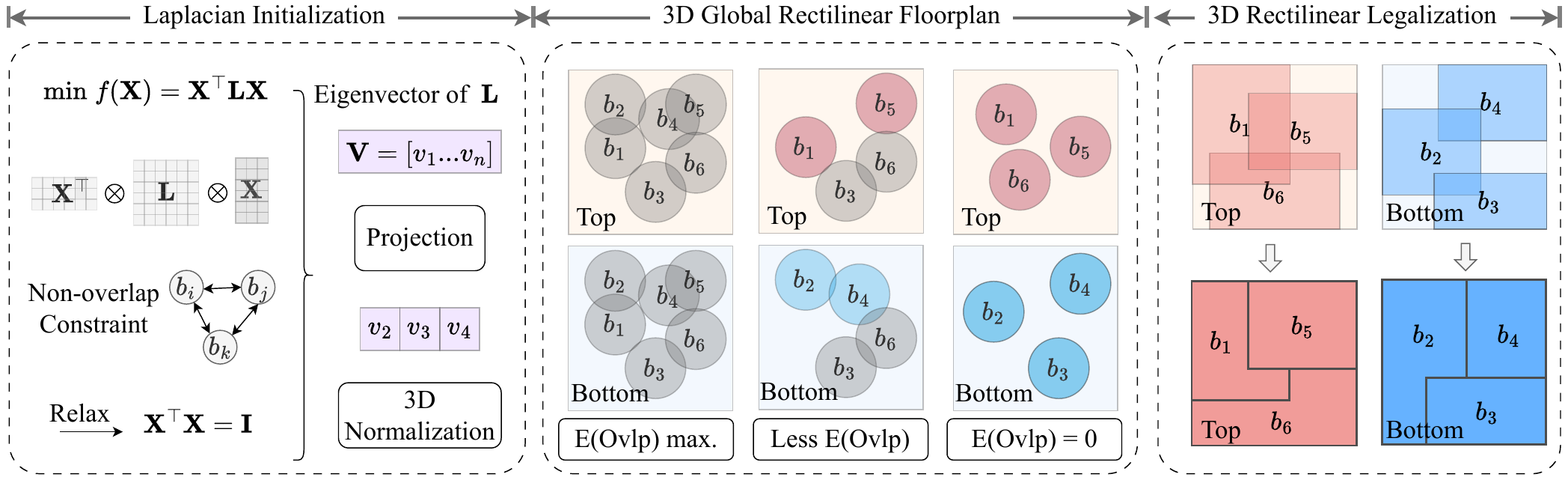}
  \caption{Overview of the proposed \textsc{Rect3D} framework for 3D-IC rectilinear floorplanning. The pipeline consists of three stages: graph Laplacian initialization (\Cref{subsec:graph_laplacian_initialization}), unified 3D global floorplanning with probabilistic die assignment (\Cref{subsec:3d_global_floorplan}), and grid-based 3D rectilinear legalization (\Cref{subsec:legalization}).}
  \label{fig:framework}
\end{figure*}
3D integration can be realized through face-to-face (F2F) hybrid bonding, through-silicon-via (TSV) based stacking, or monolithic integration~\cite{2016types-of-3d-ic, 2025ISPD-3DIC-Yu, kim2021microbumping}. In this work, we focus on F2F bonding, which naturally restricts the die set $\mathcal{T}$ to $\{\text{Top}, \text{Bottom}\}$ and avoids TSV area overhead. F2F bonding is also attractive for high-performance designs because it offers higher interconnect density and lower parasitic capacitance than TSV-based alternatives.

In floorplanning, communication cost is commonly estimated using the
half-perimeter wirelength (HPWL) model. Under two-die F2F bonding, the planar
part of this cost is captured by the 3D HPWL metric in
~\Cref{eq:hpwl3d,eq:hpwl3d_net}. The first equation aggregates the total cost
over all nets, and the second defines the planar span of a single net after
its incident blocks are projected onto their assigned dies.
\begin{equation}
\mathrm{HPWL}_{3D}(E) =  \sum_{e \in \mathcal{N}} c_e \cdot\mathrm{HPWL}_{3D}(e),
\label{eq:hpwl3d}
\end{equation}
\begin{equation}
\label{eq:hpwl3d_net}
\mathrm{HPWL}_{3D}(e) = 
\max_{b_i \in S_e}(x_i) - \min_{b_i \in S_e}(x_i)
+ \max_{b_i \in S_e}(y_i) - \min_{b_i \in S_e}(y_i),
\end{equation}
where $(x_i,y_i)$ is the center coordinate of block $b_i$ projected onto its
assigned die. The HPWL metric is widely adopted due to its simplicity and
effectiveness~\cite{2023DAC-Li,2008ASPDAC-Luo-AR}. The $z$-coordinate does not
appear explicitly in this expression because, under F2F bonding with only two
dies, vertical communication cost is represented by whether two blocks are
assigned to the same die or to opposite dies, rather than by a continuous
$z$-distance. In practice, multi-pin nets are commonly decomposed into pairwise
interactions to yield differentiable surrogate objectives during continuous
optimization~\cite{2023DAC-Li}.

\subsection{Problem Formulation}
\label{subsec:problem_formulation}

To address the coupled die-assignment and geometry challenge described in
~\Cref{sec:intro}, we formulate 3D-IC rectilinear floorplanning as the
optimization problem in~\Cref{eq:3d_formulation}. At a high level, the problem
asks for a die assignment and a rectilinear layout that jointly minimize
communication cost while satisfying geometric feasibility.

\noindent\textbf{Input}: A circuit $G = (\mathcal{B}, \mathcal{N})$ with fixed-outline dies of width $W$ and height $H$.

\noindent\textbf{Output}: 
\begin{itemize}
    \item Layer assignment $z_i \in \mathcal{T}$ for each block $b_i$;
    \item Rectilinear regions $R_i$ with center coordinates $(x_i, y_i)$ for each block $b_i$;
    \item Grid assignment matrix for grid legalization.
\end{itemize}

\noindent\textbf{Objective.}
\begin{subequations}\label{eq:3d_formulation}
\begin{alignat}{2}
    \min_{\{x_i,y_i,z_i,R_i\}} 
    & \quad \sum_{e \in \mathcal{N}} c_e \cdot \mathrm{HPWL}_{3D}(e), &\quad& \label{eq:formualtion_obj} \\
    \text{s.t. } 
    & \quad R(b_i) \cap R(b_j) = \emptyset, \quad \forall b_i \neq b_j, &\quad& \label{eq:formulation_constraint_nonoverlap} \\
    & \quad R(b_i) \in \mathcal{S}(a_i), \quad \forall b_i, &\quad& \label{eq:formulation_constraint_shape} \\
    & \quad R(b_i) \subseteq \text{Outline}, \quad \forall b_i, &\quad& \label{eq:formulation_constraint_outline_boundary}
\end{alignat}
\end{subequations}
where~\Cref{eq:formualtion_obj} minimizes the total weighted 3D HPWL;
~\Cref{eq:formulation_constraint_nonoverlap} enforces that two blocks on the
same die do not overlap;~\Cref{eq:formulation_constraint_shape} ensures that
block $b_i$ satisfies its area $a_i$ and allowable aspect-ratio set
$\mathcal{S}$; and~\Cref{eq:formulation_constraint_outline_boundary} requires
every block to stay within the fixed die boundary. Together, these constraints
make the formulation capture both electrical quality and geometric validity.

This formulation reveals three intrinsic challenges. First, die assignment and intra-die geometry are tightly coupled: a poor die decision can degrade planar wirelength and area utilization, while a poor in-die layout can invalidate an otherwise promising assignment. Second, the problem combines continuous placement variables with ultimately discrete top/bottom decisions and rectilinear-shape constraints, making direct optimization difficult. Third, the solution must remain scalable while still producing legal, connected, overlap-free rectilinear layouts. The next section presents a three-stage framework that addresses these challenges in a unified and efficient way.

\section{Methodology}
\label{sec:methodology}

To address the intrinsic challenges of 3D-IC rectilinear floorplanning formulated in~\Cref{subsec:problem_formulation}, we propose \textsc{Rect3D}, a unified and adaptive optimization framework illustrated in~\Cref{fig:framework}. The framework operates in three stages. First, graph Laplacian initialization (\Cref{subsec:graph_laplacian_initialization}) computes a topology-aware 3D seed $\mathbf{X}_0$ from netlist connectivity. Second, unified 3D global floorplanning (\Cref{subsec:3d_global_floorplan}) refines this seed into continuous block centers together with hard top/bottom die assignments. Third, grid-based rectilinear legalization (\Cref{subsec:legalization}) converts that continuous result into connected, overlap-free rectilinear regions on the assigned dies. We describe each stage in detail below.

\subsection{Graph Laplacian Initialization}
\label{subsec:graph_laplacian_initialization}
Directly solving~\Cref{eq:3d_formulation} is highly non-convex because the
solution must simultaneously determine layer assignment, intra-die coordinates,
and rectilinear block geometry under fixed-outline and non-overlap constraints.
We therefore begin with an initialization stage that takes netlist
connectivity and die scale as input and produces a topology-aware 3D seed for
the global optimizer. At this stage, we do not enforce exact rectilinear
geometry or legal non-overlap. Instead, we seek a continuous embedding that
places strongly connected blocks near each other and spreads them in 3D to
avoid degenerate layouts.

As illustrated in~\Cref{fig:topology}, we relax rectilinear regions into
circular disks while preserving adjacency relations and coarse across-die
structure. The output of this stage is a normalized coordinate matrix
$\mathbf{X}_0$, where the rescaled $z$ component acts as a soft die indicator
for the next stage.

This initialization does more than provide a better starting point. The
spectral embedding already places strongly connected blocks close to each other
and separates unrelated regions before the full penalty terms are applied. As a
result, the later global optimizer starts from a seed that already reflects the
netlist structure, making it easier to move toward a good basin of the unified
objective.

\subsubsection{Spectral Embedding Formulation}
\label{subsubsec:spectral_formulation}

To obtain a topology-preserving relaxation, we employ the graph
Laplacian~\cite{2007FOCS_Spielman_spectral}, which transforms the HPWL
objective into a spectral embedding problem. Using the connection matrix $A$
defined in~\Cref{subsec_terminology}, we first formulate the idealized problem
as
\begin{subequations}
\label{eq:3d_spectral_problem}
\begin{align}
&\min_{\mathbf{X}} \frac{1}{2} \sum_{i=1}^{n} \sum_{j=1}^{n} A_{ij} \|\mathbf{x}_{i} - \mathbf{x}_{j}\|_2^2, \label{eq:3d_objective} \\
&\text{s.t. } \quad \text{non-overlapping constraints,} \label{eq:global_floorplan_nonoverlap}
\end{align}
\end{subequations}
which penalizes squared distances between connected blocks. In other words,
blocks with large connectivity weights are encouraged to remain close in the
embedding, so the initial coordinates already reflect netlist structure.

However,~\Cref{eq:3d_spectral_problem} cannot be solved directly in spectral
form, because the non-overlapping constraint in
~\Cref{eq:global_floorplan_nonoverlap} depends on the unknown block geometries
and yields a highly non-convex feasible region. We therefore relax this strict
geometric constraint and replace it with a weaker condition that the block
embeddings remain distinct in the continuous space. This is enforced by the
orthonormality constraint~\Cref{eq:3d_orthonormal_constraints}, which prevents
all blocks from collapsing to the same location and enables an
eigendecomposition-based solution. Under this relaxation,
~\Cref{eq:3d_spectral_problem} can be simplified as
\begin{subequations}
\label{eq:native_3d_matrix_problem}
\begin{align}
&\min f_1(\mathbf{X})= \text{tr}(\mathbf{X}^\top \mathbf{L} \mathbf{X}) \label{eq:3d_matrix_objective}, \\
&\text{s.t. } \quad \mathbf{X}^\top \mathbf{X} = \mathbf{I}, \label{eq:3d_orthonormal_constraints}
\end{align}
\end{subequations}
where $\mathbf{L} = \boldsymbol{\Delta} - A$ is the graph Laplacian matrix, and
$\boldsymbol{\Delta}$ is the degree matrix with
$\Delta_{ii} = \sum_{j=1}^{n} A_{ij}$. This matrix form makes the relaxed
problem directly solvable by eigendecomposition and exposes the smoothest
nontrivial modes of the connectivity graph.
\begin{figure}[t!]
    \centering
    \includegraphics[width=\linewidth]{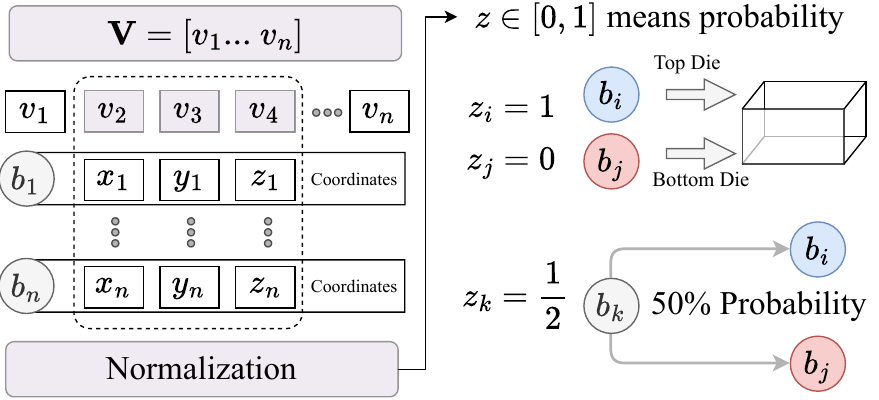}
    \caption{Projection and normalization of eigenvectors. The raw 3D spectral coordinates are affinely mapped to the die outline in the $(x,y)$-plane and rescaled to $[0,1]$ along $z$.}
    \label{fig:projection}
\end{figure}
\subsubsection{Eigenvector Selection and Coordinate Assignment}
\label{subsubsec:eigenvector_selection}

After computing the eigendecomposition of $\mathbf{L}$, we discard the trivial
constant eigenvector $\mathbf{v}_1$ (corresponding to eigenvalue
$\lambda_1=0$) implied by~\Cref{eq:3d_orthonormal_constraints}, and select the
next three eigenvectors to initialize block coordinates as
~\Cref{eq:3d_coordinate_init}. These three vectors provide the raw 3D
coordinates before normalization.
\begin{equation}
\label{eq:3d_coordinate_init}
\mathbf{X} = [\mathbf{v}_2, \mathbf{v}_3, \mathbf{v}_4] \in \mathbb{R}^{n \times 3}
\end{equation}
where $\mathbf{v}_2,\mathbf{v}_3,\mathbf{v}_4$ correspond to the three smallest
nonzero eigenvalues $\lambda_2 \le \lambda_3 \le \lambda_4$ of $\mathbf{L}$. This choice is the minimizer of
\Cref{eq:3d_matrix_objective} under the orthonormality constraint and is unique
up to an orthogonal transformation~\cite{2007FOCS_Spielman_spectral}.

A natural question is why we assign $\mathbf{v}_2$ and $\mathbf{v}_3$ to the
$(x,y)$ coordinates and $\mathbf{v}_4$ to the $z$-coordinate. The Laplacian
eigenvectors are ordered by their associated eigenvalues, and lower
eigenvalues correspond to smoother partitions of the netlist graph. The
Fiedler vector $\mathbf{v}_2$, the eigenvector with the smallest nonzero
eigenvalue, captures the most significant bipartition of the connectivity
structure, and $\mathbf{v}_3$ captures the next most significant one.
Together, $\mathbf{v}_2$ and $\mathbf{v}_3$ provide the most informative planar
embedding and are therefore assigned to the $x$ and $y$ coordinates, where the
spatial resolution is highest. The third eigenvector $\mathbf{v}_4$ provides a
coarser partition that naturally corresponds to the binary die assignment
along the $z$-axis, since the $z$-dimension ultimately reduces to a top/bottom
decision. Any permutation of these three eigenvectors only applies a rigid
rotation or reflection to the 3D embedding and does not change the Laplacian
energy. In practice, however, assigning the smoothest eigenvectors to the
planar dimensions and the next one to the vertical dimension yields the best
initialization, as confirmed by our ablation studies.

\subsubsection{Affine Normalization}
\label{subsubsec:affine_normalization}

The embedding is then standardized and affinely mapped to the die outline
according to~\Cref{eq:coordinate_standardization}. This step fixes the scale
and location of the raw spectral coordinates so that the initialization can be
used directly by the global floorplanner.
\begin{equation}
\label{eq:coordinate_standardization}
\mathbf{X}_0 = (\mathbf{X} - \mathbf{1}\boldsymbol{\mu}^\top)\boldsymbol{\Sigma}^{-1}\mathbf{S} + \mathbf{1}\mathbf{t}^\top
\end{equation}
where $\boldsymbol{\mu} \in \mathbb{R}^3$ and $\boldsymbol{\Sigma} \in
\mathbb{R}^{3 \times 3}$ (diagonal) denote the per-dimension mean and standard
deviation of $\mathbf{X}$, respectively, while $\mathbf{S} \in \mathbb{R}^{3
\times 3}$ and $\mathbf{t} \in \mathbb{R}^3$ are the scaling and translation
parameters that map the normalized coordinates to the die outline dimensions.
Concretely, the $x$ and $y$ coordinates are mapped to the die outline
$[0, W] \times [0, H]$, while the $z$ coordinates are monotonically rescaled
into $[0,1]$. These rescaled $z$ coordinates are subsequently interpreted as
soft die indicators, so that values close to~0 or~1 correspond to a strong
preference for the bottom or top die, respectively. This
projection-and-normalization step is illustrated in~\Cref{fig:projection}.

In this way, the Laplacian embedding provides a 3D continuous initialization
that already encodes meaningful cross-die structure. It is not yet a legal
floorplan. Rather, it yields the normalized seed $\mathbf{X}_0$, whose planar
coordinates initialize block centers and whose rescaled $z$ values serve as
soft die indicators for the unified global floorplanning stage described next.

\subsection{Unified 3D Global Floorplanning}
\label{subsec:3d_global_floorplan}

This stage takes the normalized seed $\mathbf{X}_0$ from
\Cref{subsec:graph_laplacian_initialization} and jointly refines the planar
block centers and the soft die variables. Its output is a continuous floorplan
with hard top/bottom die assignments, which is then passed to the legalization
stage. At a high level, the Stage-2 flow summarized in
~\Cref{alg:lbfgsb_floorplan} repeatedly evaluates the unified objective under
the current regularization weight, computes a projected L-BFGS-B update, and
advances the solution until convergence. Across the three scheduled phases, the
$z$-variables evolve from soft die probabilities to hard top/bottom
assignments. We next define the objective and explain how adaptive
regularization and L-BFGS-B realize this process.

\subsubsection{Unified Objective Function}
\label{subsubsec:unified_objective}

The unified 3D objective is defined in~\Cref{eq:unified_3d_objective}.
\begin{equation}
\label{eq:unified_3d_objective}
f(\mathbf{X}) = f_{\text{HPWL}}(\mathbf{X}) + \sum_{c \in \mathcal{C}} \lambda_c g_c(\mathbf{X}) + \alpha(t) R(\mathbf{z}),
\end{equation}
where $f_{\text{HPWL}}(\mathbf{X})$ minimizes wirelength, the penalty terms $g_c(\mathbf{X})$ enforce physical feasibility, and the regularizer $R(\mathbf{z})$ gradually drives the probabilistic die variables toward discrete assignments.
We model the wirelength objective using the standard pairwise decomposition~\cite{2023DAC-Li}, as shown in~\Cref{eq:pairwise_hpwl}.
\begin{equation}
\label{eq:pairwise_hpwl}
f_{\text{HPWL}}(\mathbf{X}) = \sum_{(i,j)} A_{ij} \bigl(|x_i - x_j| + |y_i - y_j|\bigr),
\end{equation}
where $A_{ij}$ is the connection weight between blocks $b_i$ and $b_j$.
Although the absolute-value function is non-smooth at the origin, this poses no
practical difficulty because the subsequent L-BFGS-B optimization approximates
gradients by numerical finite differences, which naturally handles such points.

\algrenewcommand{\algorithmiccomment}[1]{\hfill{$\triangleright$\,#1}}

\newcommand{\eqnum}[1]{(\ref{#1})}
\newcommand{\eqrange}[2]{(\ref{#1}--\ref{#2})}

\begin{algorithm}[t!]
\caption{L-BFGS-B for 3D rectilinear floorplanning.}
\small
\label{alg:lbfgsb_floorplan}
\begin{algorithmic}[1]
\Require initialized $\mathbf{X}_{0}$ from \eqrange{eq:3d_coordinate_init}{eq:coordinate_standardization}
\Ensure  3D global floorplan $\mathbf{X}^{*}$
\State $\mathbf{X}\leftarrow\mathbf{X}^{(0)}$;\; $\mathcal{M}\leftarrow\varnothing$
\For{$k=1,\dots,K$}
  \State $\alpha\leftarrow\alpha(k)$ \Comment{update probability schedule}
  \State $f\leftarrow f(\mathbf{X};\alpha)$ \Comment{add 3D probabilities}
  \State $g\leftarrow \nabla f(\mathbf{X};\alpha)$ \Comment{see \Cref{eq:unified_3d_objective}}
  \State $\mathbf{p}\leftarrow \textsc{GetDirection}(\mathcal{M},g)$ \Comment{see \eqnum{eq:two_loop}}
  \State $\eta\leftarrow\textsc{LineSearch}(\mathbf{X},\mathbf{p})$ \Comment{see \eqnum{eq:armijo_condition}}
  \State $\mathbf{X}_{\text{new}}\leftarrow \textsc{ProjectBox}(\mathbf{X}+\eta\mathbf{p})$ \Comment{$z\in[0,1]$}
  \State $s_k\leftarrow\mathbf{X}_{\text{new}}-\mathbf{X}$ \Comment{see \eqnum{eq:lbfgs_s_y_pair}}
  \State $g_{\text{new}}\leftarrow\nabla f(\mathbf{X}_{\text{new}};\alpha)$
  \State $y_k\leftarrow g_{\text{new}}-g$ \Comment{see \eqnum{eq:lbfgs_s_y_pair}}
  \State $\mathcal{M}\leftarrow\textsc{UpdateMemory}(\mathcal{M},(s_k,y_k),m)$
  \State $\mathbf{X}\leftarrow\mathbf{X}_{\text{new}}$
  \If{$\|\nabla f(\mathbf{X})\|_\infty \le \varepsilon_g$ \textbf{ or } rel.\ dec.\ $\le \varepsilon_f$}
     \State \textbf{break}
  \EndIf
\EndFor
\State $\mathbf{X}^{*} \leftarrow \mathbf{X}$
\end{algorithmic}
\end{algorithm}

The constraint set $\mathcal{C} = \{\text{overlap}, \text{outline},
\text{balance}\}$ enforces the physical feasibility conditions in
~\Cref{eq:formulation_constraint_nonoverlap,eq:formulation_constraint_outline_boundary}.
These three penalties play distinct roles: the overlap term discourages
same-die conflicts, the outline term keeps blocks within the die boundary, and
the balance term prevents the optimization from collapsing almost all area onto
one die. Since blocks carry probabilistic die assignments during optimization,
the overlap term is defined in expectation by~\Cref{eq:overlap_penalty}.
\begin{equation}
\label{eq:overlap_penalty}
g_{\text{overlap}}(\mathbf{X}) = \sum_{i < j} P_{ij} \cdot \max\!\left(0,\; r_i + r_j - d_{ij}\right)^2,
\end{equation}
where $d_{ij} = \|\mathbf{x}_i - \mathbf{x}_j\|_2$ is the Euclidean distance
between block centers projected onto the $(x,y)$-plane, $r_i$ and $r_j$ are
effective radii derived from block areas, and $P_{ij}$ is the probability that
blocks $b_i$ and $b_j$ are assigned to the same die. The same-layer
probability is computed by~\Cref{eq:same_layer_prob}.
\begin{equation}
\label{eq:same_layer_prob}
P_{ij} = z_i z_j + (1 - z_i)(1 - z_j),
\end{equation}
which equals~1 when both blocks are on the same die ($z_i = z_j \in \{0,1\}$),
equals~0 when they are assigned to opposite dies ($z_i,z_j \in \{(0,1),(1,0)\}$),
and becomes~0.5 when both assignments are maximally ambiguous
($z_i = z_j = 0.5$). In this way, overlap penalties are weighted by the
likelihood of co-location, allowing blocks assigned to different dies to
overlap freely in the $(x,y)$-plane during early exploration.

To keep blocks inside the die outline, we add the boundary penalty in~\Cref{eq:boundary_penalty,eq:outline_component}.
\begin{equation}
\label{eq:boundary_penalty}
g_{\text{outline}}(\mathbf{X}) = \sum_{i=1}^{n} \bigl[h(x_i; W, r_i) + h(y_i; H, r_i)\bigr],
\end{equation}
\begin{equation}
\label{eq:outline_component}
h(u; U, r) = \max(0, u - U + r)^2 + \max(0, -u + r)^2,
\end{equation}
where $W$ and $H$ are the die width and height. This quadratic term vanishes
when all blocks lie inside the outline and increases smoothly as any block
protrudes beyond the boundary. To avoid degenerate solutions in which nearly
all block area collapses onto one die, we further introduce the layer-balance
term in~\Cref{eq:balance_penalty}.
\begin{equation}
\label{eq:balance_penalty}
g_{\text{balance}}(\mathbf{X}) = \left(\frac{\sum_{i=1}^{n} z_i \cdot a_i}{\sum_{i=1}^{n} a_i} - \rho\right)^2,
\end{equation}
where $\rho$ is the target area ratio for the top die (set to 0.5 for balanced allocation). This penalty encourages an even distribution of total block area across both dies.

The penalty weights $\lambda_{\text{overlap}}$, $\lambda_{\text{outline}}$, and $\lambda_{\text{balance}}$ are set to fixed values that are large enough to enforce constraint satisfaction throughout the optimization.

To progressively convert the relaxed die variables into discrete assignments, we use the regularization term in~\Cref{eq:z_regularization}.
\begin{equation}
\label{eq:z_regularization}
R(\mathbf{z}) = \sum_{i=1}^n 4z_i(1-z_i).
\end{equation}
The function~\Cref{eq:z_regularization} reaches its maximum at $z_i = 0.5$ and
vanishes when $z_i \in \{0,1\}$, so it penalizes ambiguous die assignments.
Its weight $\alpha(t)$ follows a three-phase schedule from $\alpha_{\min}$ to
$\alpha_{\text{mid}}$ and then to $\alpha_{\max}$, gradually pushing the
$z$-values from soft probabilities to binary die assignments. Each phase runs
L-BFGS-B to convergence before the next phase begins. During Phases~1 and~2,
overlap is evaluated in expectation using~\Cref{eq:overlap_penalty,eq:same_layer_prob}. In
Phase~3, each block is hard-assigned according to whether $z_i > 0.5$, and
overlap is evaluated only within the same die.

We solve the resulting box-constrained problem using
L-BFGS-B~\cite{lbfgsb_proposed}, a limited-memory quasi-Newton method that
naturally handles the bound constraints $z_i \in [0,1]$. For readers who do
not wish to follow the solver details, the main point is that each iteration
updates the continuous coordinates and die probabilities while respecting box
bounds, and the three-phase schedule changes only the regularization weight.
In our implementation, gradients are approximated by L-BFGS-B's built-in
numerical finite-difference scheme. This avoids hand-derived analytic
gradients and also handles the non-smooth absolute values in the HPWL term
~\Cref{eq:pairwise_hpwl} without extra machinery. The algorithm maintains a
compact approximation $H_k$ of the inverse Hessian using the $m$ most recent
curvature pairs in~\Cref{eq:lbfgs_s_y_pair,eq:two_loop,eq:armijo_condition}.
\begin{equation}
\label{eq:lbfgs_s_y_pair}
s_k = \mathbf{X}_{k+1} - \mathbf{X}_{k}, \quad
y_k = \nabla f(\mathbf{X}_{k+1}) - \nabla f(\mathbf{X}_{k}).
\end{equation}
\begin{equation}
\label{eq:two_loop}
\mathbf{p} = -H_k \mathbf{g}, \quad H_k = \text{L-BFGS}(\{s_i, y_i\}_{i=k-m}^{k-1}, H_0),
\end{equation}
\begin{equation}
\label{eq:armijo_condition}
f(\mathbf{X} + \eta \mathbf{p})
\leq f(\mathbf{X}) + c \eta \nabla f(\mathbf{X})^\top \mathbf{p}, \quad c \in (0,1).
\end{equation}
In this update,~\Cref{eq:two_loop} determines the search direction from the
recent curvature history, while~\Cref{eq:armijo_condition} determines the step
size through backtracking line search. The initial scaling is set to
$H_0 = \frac{s_{k-1}^\top y_{k-1}}{y_{k-1}^\top y_{k-1}} \mathbf{I}$ from the
most recent curvature pair. This limited-memory update avoids forming the full
$3n \times 3n$ Hessian and requires only $\mathcal{O}(mn)$ storage. The updated
iterate is then projected onto the box constraints:
$\mathbf{X}_{\text{new}} = \Pi_{[\mathbf{l},\mathbf{u}]}(\mathbf{X} + \eta \mathbf{p})$,
where $\Pi$ denotes the componentwise projection onto the box $[\mathbf{l}, \mathbf{u}]$, with $l_i = 0, u_i = 1$ for the $z$-coordinates and $l_i = 0, u_i = W$ (or $H$) for the $x$ (or $y$) coordinates.

In~\Cref{alg:lbfgsb_floorplan}, lines~3--8 perform one projected L-BFGS-B
update under the current scheduled weight $\alpha(k)$, and lines~9--15 update
the curvature history and test convergence. This loop is repeated within each
phase until convergence.
\begin{algorithm}[t]
\caption{Grid-based 3D rectilinear legalization.}
\small
\label{alg:rect_legalization}
\begin{algorithmic}[1]
\Require centers $\{(x_i^*,y_i^*,z_i^*)\}_{i=1}^{n_b}$, areas $\{a_i\}$
\Ensure  binary assignment $M \in \{0,1\}^{n_b \times |\mathcal{G}|}$
\State construct grid $\mathcal{G}$ with $m \times n$ cells
\State compute $a_{\text{cell}}$ and $n_i^{\text{target}}$; see \Cref{eq:grid_cell_area,eq:target_cell_count}
\For{each die $d \in \{\text{Top},\text{Bottom}\}$}
  \State $\mathcal{B}_d \leftarrow \{\, i \mid z_i^* \text{ on die } d \,\}$ \Comment{blocks on die $d$}
  \For{each cell $g \in \mathcal{G}$}
    \State nearest-center assignment; see \Cref{eq:legalization_3d,eq:legal_cell_cost}
  \EndFor
  \For{$k = 1,\dots,K_{\text{CA}}$}
    \State compute counts $c_i \leftarrow \sum_{g} M_{i,g}$ for all $i \in \mathcal{B}_d$
    \State changed $\leftarrow$ \textbf{false}
    \For{each cell $g \in \mathcal{G}$}
      \State update the label of $g$ by the CA rule; see \Cref{eq:ca_deficit}
      \State update $c_i$; set changed $\leftarrow$ \textbf{true} if needed
    \EndFor
    \If{\textbf{not} changed}
      \State \textbf{break}
    \EndIf
  \EndFor
\EndFor
\State \Return $M$
\end{algorithmic}
\end{algorithm}

The resulting Stage-2 solution corresponds to the middle panel of
~\Cref{fig:framework}: continuous block centers together with hard top/bottom
die assignments, ready for Stage~3. This probabilistic-to-discrete evolution
is further analyzed in the case study in~\Cref{subsec:flow_analysis}.

\subsection{Rectilinear Legalization}
\label{subsec:legalization}

\begin{figure}[t]
    \centering
    \includegraphics[width=.9\linewidth]{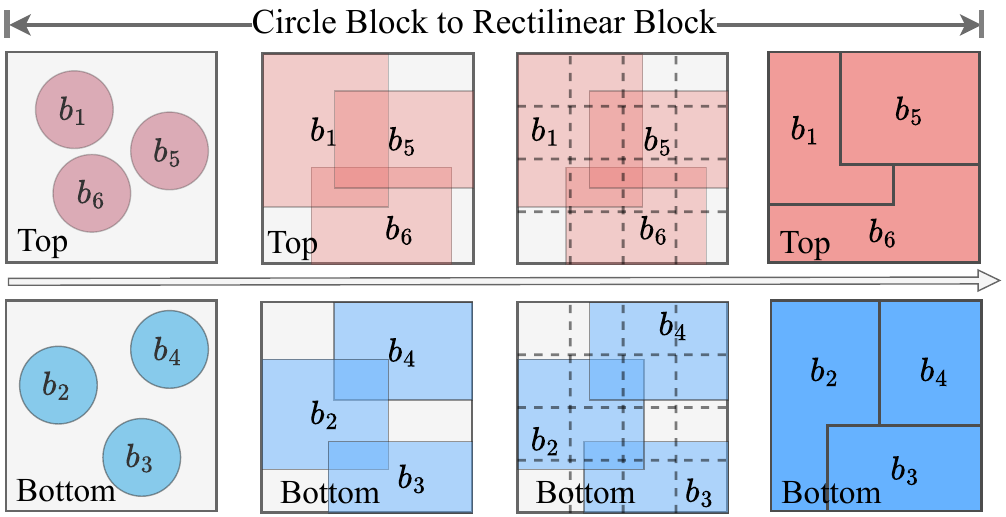}
    \caption{Illustration of the procedure transforming circular blocks into rectilinear blocks on the top and bottom dies. Continuous circular embeddings are discretized onto a grid, where each block occupies a connected cluster of grid cells forming a rectilinear polygon.}
    \label{fig:rect_legalization}
\end{figure}

Building on the unified 3D global floorplanning stage in
\Cref{subsec:3d_global_floorplan}, which outputs continuous circular embeddings
$\mathbf{X}^*$ with binary layer assignments after Phase~3 convergence, this
stage converts the continuous solution into a discrete 3D rectilinear
floorplan. Once die assignment is fixed, legalization is carried out on a
shared-outline 2D grid for each die and then combined into the final 3D
result. As shown in~\Cref{fig:rect_legalization}, circular blocks on each die
are snapped to a grid and expanded into unions of grid cells whose
staircase-like boundaries form rectilinear polygons.
\Cref{alg:rect_legalization} summarizes the overall procedure.

We first discretize the shared die outline into a two-dimensional grid
$\mathcal{G}$ with $m$ rows and $n$ columns, following
\Cref{alg:rect_legalization}, lines~1--2. For face-to-face bonding with a shared
outline, the 3D grid is obtained by stacking two identical copies of
$\mathcal{G}$ along the $z$-axis, one for each die. The grid-cell area and the
target cell count of each block are defined by
\begin{align}
\label{eq:grid_cell_area}
a_{\text{cell}} &= (W\cdot H)/(mn), \\
\label{eq:target_cell_count}
n_i^{\text{target}} &= \operatorname{round}(a_i / a_{\text{cell}}).
\end{align}
Here, $W$ and $H$ are the die width and height. In practice,
we choose $m$ and $n$ according to the benchmark size so that the grid is fine
enough to approximate target areas while keeping the legalization manageable.
Finer grids generally provide a better approximation of target areas and
produce smoother rectilinear boundaries.

Given this grid, legalization can be viewed as a discrete assignment problem.
At a high level, the goal is to replace each continuous circular block by a
connected set of grid cells whose area matches the target block area and whose
location stays close to the continuous Stage-2 solution.
Let $n_b = |\mathcal{B}|$ be the number of blocks and $|\mathcal{G}|$ the number
of grid cells per die. We introduce a binary assignment matrix
$M \in \{0,1\}^{n_b \times |\mathcal{G}|}$, where $M_{i,g}=1$ indicates that grid
cell $g \in \mathcal{G}$ is assigned to block $b_i$ on its designated die. The
objective is formulated in~\Cref{eq:legalization_3d}.
\begin{equation}
\label{eq:legalization_3d}
\min_{M \in \{0,1\}^{n_b \times |\mathcal{G}|}} 
\;\; \sum_{i=1}^{n_b} \sum_{g \in \mathcal{G}} c_{i,g}\, M_{i,g},
\end{equation}
where $c_{i,g}$ is a per-cell cost that penalizes deviation from the continuous
centroid of $b_i$. Thus, the objective prefers grid cells near the continuous
location of each block while still allowing the final region to be assembled
cell by cell. Specifically,
\begin{equation}
\label{eq:legal_cell_cost}
c_{i,g} = \left\|\mathbf{c}(g) - \pi(x_i^*,y_i^*)\right\|_1,
\end{equation}
where $\pi(x_i^*,y_i^*)$ denotes the nearest grid point to $(x_i^*,y_i^*)$, so
that cells closer to the continuous solution are preferred.
The assignment is subject to the constraints in~\Cref{eq:legal_single_occupancy,eq:legal_area_preservation}.
\begin{alignat}{2}
&\sum_{i=1}^{n_b} M_{i,g} \le 1, &\quad& \forall g \in \mathcal{G}, \label{eq:legal_single_occupancy} \\
&\sum_{g \in \mathcal{G}} M_{i,g} = n_i^{\text{target}}, &\quad& \forall b_i \in \mathcal{B}, \label{eq:legal_area_preservation}
\end{alignat}
where~\Cref{eq:legal_single_occupancy} enforces single-cell occupancy
(non-overlap) and~\Cref{eq:legal_area_preservation} preserves the target area
of each block. These two constraints express the core legalization
requirements: no cell can be shared by multiple blocks, and each block must
receive the correct number of cells.

Directly solving~\Cref{eq:legalization_3d} as a large-scale integer linear
program would be computationally expensive on the grid sizes of interest. We
therefore adopt the heuristic procedure summarized in
~\Cref{alg:rect_legalization}, lines~3--31. For each die, the
algorithm first performs a nearest-centroid assignment of all grid cells
(lines~5--7), which yields a Voronoi-style initialization aligned with the
continuous coordinates $\mathbf{X}^*$. This initialization is geometrically
natural and already gives each block a connected region centered around its
continuous position.
The subsequent refinement is carried out by a cellular automaton over
lines~8--30. In each sweep, the current cell counts
$c_i = \sum_g M_{i,g}$ are first computed for all blocks (line~9), after which
the grid cells are revisited in raster order (line~11). The label update uses
the 4-neighborhood of each cell together with the deficit
\begin{equation}
\label{eq:ca_deficit}
\delta_i = n_i^{\text{target}} - c_i.
\end{equation}
Over-filled blocks may release boundary cells to under-filled neighboring
blocks, while unassigned cells adopt nearby labels in a way that favors
spatial coherence and area correction. Crucially, any candidate reassignment
is accepted only when the donor block remains connected. This connectivity
check preserves contiguity throughout refinement and prevents fragmentation
into disconnected pieces.

The CA sweep is repeated until no label changes occur or a maximum iteration
count $K_{\text{CA}}$ is reached, as indicated in
\Cref{alg:rect_legalization}, lines~18--30. In practice, convergence occurs
within a small number of iterations, giving a per-sweep time complexity of
$\mathcal{O}(n_b\,|\mathcal{G}|)$.

Together, the Voronoi-style initialization and the connectivity-preserving CA
updates ensure that the final assignment $M$ produces a connected rectilinear
region for every block on its assigned die. The staircase boundaries of these
grid-cell clusters naturally form rectilinear (orthogonal) polygons,
satisfying the shape constraint in~\Cref{eq:formulation_constraint_shape}.

The output of this stage is the final legal 3D rectilinear floorplan: for each
block $b_i$, a connected rectilinear region placed on the assigned die and
satisfying the constraints in~\Cref{eq:3d_formulation}.

\section{Experimental Results}
\label{sec:experiments}

\begin{table*}[t!]
    \centering
    \footnotesize
    \tabcolsep=3.5pt
    \def\arraystretch{1.2}
    \caption{Experimental results on GSRC benchmarks~\cite{GSRC-bench} with 3D rectangular floorplan baseline methods.}
    \label{tab:comparison_methods_doubled_runT}
    \begin{threeparttable}
    \resizebox{\linewidth}{!}{%
    \begin{tabular}{c|cc|cc|cc|cc|cc|cc|cc}
    \toprule
    \multirow{2}{*}{\textbf{Case}}
      & \multicolumn{2}{c|}{\textbf{FM\cite{2022TCAD-snap3d}+AR\cite{2008ASPDAC-Luo-AR}}}
      & \multicolumn{2}{c|}{\textbf{FM+DAC23\cite{2023DAC-Li}}}
      & \multicolumn{2}{c|}{\textbf{Spectral\cite{1995DAC-Spectral_partitioning}+AR}}
      & \multicolumn{2}{c|}{\textbf{Spectral+DAC23}}
      & \multicolumn{2}{c|}{\textbf{GSP\cite{2011TVLSI-grouped-sequence-pair}}}
      & \multicolumn{2}{c|}{\textbf{Lin\cite{2021TVLSI-thermal-aware-3dfloorplan-tsv}}}
      & \multicolumn{2}{c}{\textbf{\textsc{Rect3D}} (ours)} \\
    \cmidrule(lr){2-3}
    \cmidrule(lr){4-5}
    \cmidrule(lr){6-7}
    \cmidrule(lr){8-9}
    \cmidrule(lr){10-11}
    \cmidrule(lr){12-13}
    \cmidrule(lr){14-15}
    & \textbf{WL} & \textbf{Time}
    & \textbf{WL} & \textbf{Time}
    & \textbf{WL} & \textbf{Time}
    & \textbf{WL} & \textbf{Time}
    & \textbf{WL} & \textbf{Time}
    & \textbf{WL} & \textbf{Time}
    & \textbf{WL (±Std)} & \textbf{Time} \\
    \midrule
    n10
      & 43223 & 6.72
      & 48881 & 2.16
      & 24571 & 5.04
      & 28020 & 2.08
      & 36444 & 7.62
      & 41166 & 41.1
      & 12090 (±0.02) & 9.5 \\
    n30
      & 84456 & 13.84
      & 88703 & 5.68
      & 67193 & 15.68
      & 72048 & 10.00
      & 100818 & 41.86
      & 79429 & 55.76
      & 32782 (±631) & 22.5 \\
    n50
      & 217731 & 21.44
      & 219508 & 18.32
      & 144133 & 32.56
      & 147835 & 18.82
      & 174368 & 109.74
      & 203847 & 159.06
      & 73418 (±496) & 28.0 \\
    n100
      & 402324 & 37.04
      & 388336 & 102.16
      & 268341 & 45.36
      & 259307 & 107.84
      & 385398 & 351.9
      & 373565 & 161.94
      & 118976 (±10211) & 34.3 \\
    n200
      & 880486 & 130.16
      & 852036 & 1074.32
      & 446443 & 115.04
      & 408192 & 1105.04
      & 397594 & 1360.86
      & 832867 & 772.84
      & 258198 (±25387) & 37.5 \\
    n300
      & 1195266   & 401.13
      & -- & --
      & 505679  & 537.6
      & -- & --
      & 9499655 & 1800
      & 1139620   & 893
      & 405424.89 (±24857) & 42.1 \\
    \midrule
    Avg. ratio
      & 3.143 & 2.695
      & 3.260 & 6.552
      & 1.879 & 3.258
      & 2.058 & 6.789
      & 6.113 & 15.981
      & 2.963 & 9.838
      & \textbf{1} & \textbf{1} \\
    \bottomrule
    \end{tabular}
    }
    {\footnotesize
    \par\vspace{3pt}
    \noindent\parbox{\linewidth}{\textit{*} Avg. ratio is computed as the arithmetic mean of WL (or Time) relative to \textsc{Rect3D} over all available cases from n10 to n300. For methods that do not finish n300 within the 1800s time limit, the average is taken over n10--n200 only. WL is measured in \si{\micro\meter} and Time in seconds. For clarity of presentation, baseline methods are reported with averaged results only, whereas our \textsc{Rect3D} further includes the mean and standard deviation over five runs.}
    }
    \end{threeparttable}
\end{table*}

\begin{table*}[t!]
    \centering
    \footnotesize
    \tabcolsep=3.5pt
    \def\arraystretch{1.2}
    \caption{Experimental results on GSRC benchmarks with 3D rectilinear floorplan baselines using the same Rect3D legalization.}
    \label{tab:comparison_rectilinear}
    \begin{threeparttable}
    \resizebox{\linewidth}{!}{%
    \begin{tabular}{c|cc|cc|cc|cc|cc|cc|cc}
    \toprule
    \multirow{2}{*}{\textbf{Case}}
      & \multicolumn{2}{c|}{\textbf{FM+ISEDA\cite{2025ISEDA_Rectilinear}}}
      & \multicolumn{2}{c|}{\textbf{FM+Jigsaw\cite{2024ICCAD_Jisaw}}}
      & \multicolumn{2}{c|}{\textbf{FM+Modern\cite{2024ICCAD_Chen_Rectilinear_soft}}}
      & \multicolumn{2}{c|}{\textbf{Spectral+ISEDA}}
      & \multicolumn{2}{c|}{\textbf{Spectral+Jigsaw}}
      & \multicolumn{2}{c|}{\textbf{Spectral+Modern}}
      & \multicolumn{2}{c}{\textbf{\textsc{Rect3D}} (ours)} \\
    \cmidrule(lr){2-3}
    \cmidrule(lr){4-5}
    \cmidrule(lr){6-7}
    \cmidrule(lr){8-9}
    \cmidrule(lr){10-11}
    \cmidrule(lr){12-13}
    \cmidrule(lr){14-15}
    & \textbf{WL} & \textbf{Time}
    & \textbf{WL} & \textbf{Time}
    & \textbf{WL} & \textbf{Time}
    & \textbf{WL} & \textbf{Time}
    & \textbf{WL} & \textbf{Time}
    & \textbf{WL} & \textbf{Time}
    & \textbf{WL (±Std)} & \textbf{Time} \\
    \midrule
    n10
      & 49982 & 2.99
      & 49982 & 2.88
      & 55944 & 0.13
      & 34736 & 3.00
      & 34736 & 3.02
      & 38088 & 0.37
      & 12090 (±0.02) & 9.5 \\
    n30
      & 125439 & 4.49
      & 125438 & 4.29
      & 137170 & 0.93
      & 97909 & 4.79
      & 97909 & 4.77
      & 100014 & 1.00
      & 32782 (±631) & 22.5 \\
    n50
      & 262307 & 6.66
      & 261963 & 6.49
      & 266647 & 2.29
      & 185156 & 7.39
      & 186321 & 7.36
      & 208583 & 1.92
      & 73418 (±496) & 28.0 \\
    n100
      & 489486 & 10.33
      & 489242 & 10.17
      & 499904 & 4.87
      & 293453 & 11.68
      & 294647 & 11.53
      & 354449 & 4.94
      & 118976 (±10211) & 34.3 \\
    n200
      & 847725 & 18.43
      & 848560 & 18.04
      & 1073570 & 16.04
      & 415823 & 24.55
      & 415481 & 24.54
      & 620344 & 16.37
      & 258198 (±25387) & 37.5 \\
    n300
      & 1727701 & 23.98
      & 1727800 & 23.81
      & 1553847 & 31.86
      & 789731 & 32.45
      & 789941 & 32.25
      & 818068 & 30.86
      & 405425 (±24857) & 42.1 \\
    \midrule
    Avg. ratio
      & 3.865 & 0.352
      & 3.865 & 0.345
      & 4.106 & 0.244
      & 2.401 & 0.426
      & 2.405 & 0.425
      & 2.740 & 0.244
      & \textbf{1} & \textbf{1} \\
    \bottomrule
    \end{tabular}
    }
    {\footnotesize
    \par\vspace{3pt}
    \noindent\parbox{\linewidth}{\textit{*} Each baseline uses a partition-first flow: FM~\cite{2022TCAD-snap3d} or Spectral~\cite{1995DAC-Spectral_partitioning}, per-die 2D global optimization, and Rect3D legalization. Avg. ratio is the arithmetic mean relative to \textsc{Rect3D} over n10--n300. WL is in \si{\micro\meter} and Time in seconds. Baseline WL is deterministic across runs; only runtime is averaged. Time ratios below 1 mean the baseline is faster than \textsc{Rect3D}.}
    }
    \end{threeparttable}
\end{table*}

The proposed method was implemented in Python 3.11.
All experiments were conducted on a Linux workstation running Ubuntu 22.04 LTS, equipped with dual Intel Xeon Gold 6426Y processors and 256\,GB RAM. We evaluate all GSRC benchmarks~\cite{GSRC-bench} using their soft-block netlists.
Following common practice in floorplanning evaluation~\cite{2023DAC-Li,2008ASPDAC-Luo-AR}, fixed I/O pads and their incident nets are excluded from the optimization.

\subsection{Comparison with 3D Rectangular Floorplanning Baselines}
\label{subsec:cmp_with_baselines}
\noindent
We first compare \textsc{Rect3D} against representative 3D rectangular
floorplanning baselines. Across all valid cases, \textsc{Rect3D} achieves the
best wirelength, and it also delivers the best average runtime over the
benchmark suite. The six baselines
span both partition-first and combinatorial methods. Specifically, the partition-first baselines are (1)~FM+AR, (2)~FM+DAC23, (3)~Spectral+AR, and (4)~Spectral+DAC23, which combine FM partitioning~\cite{1995ICCAD-FM_partition} or spectral partitioning~\cite{1995DAC-Spectral_partitioning} with AR~\cite{2008ASPDAC-Luo-AR} or DAC23~\cite{2023DAC-Li}; the combinatorial baselines are (5)~GSP~\cite{2011TVLSI-grouped-sequence-pair} and (6)~Lin~\cite{2021TVLSI-thermal-aware-3dfloorplan-tsv}.
This baseline set is intentionally diverse: if \textsc{Rect3D} consistently
outperforms both families, then its advantage cannot be attributed to weakness
in one particular competing flow. \Cref{tab:comparison_methods_doubled_runT}
summarizes wirelength (WL) and runtime across all GSRC instances.
\textsc{Rect3D} is the only method that achieves the lowest WL on every
instance while still delivering the best average runtime overall. Its runtime
is not always the smallest on the easiest benchmarks, but it becomes the
fastest on the larger cases.
Aggregating over all available cases, \textsc{Rect3D} reduces wirelength by
68.2\% vs.\ FM+AR ($3.14\times$ smaller),
69.3\% vs.\ FM+DAC23 ($3.26\times$ smaller),
46.8\% vs.\ Spectral+AR ($1.88\times$ smaller),
51.4\% vs.\ Spectral+DAC23 ($2.06\times$ smaller),
83.6\% vs.\ GSP ($6.11\times$ smaller),
and 66.3\% vs.\ Lin ($2.96\times$ smaller).

The consistent advantage across all six baselines shows that the unified
optimization of die assignment and layout geometry in \textsc{Rect3D} yields
better solutions than either partition-first or combinatorial approaches. In
particular, the large gap relative to FM+AR and FM+DAC23 highlights the cost
of fixing die assignment before layout optimization: even a strong 2D
floorplanner often struggles to recover from a suboptimal partition. Even the strongest
spectral variants remain about $1.9\times$--$2.1\times$ worse in average WL,
which indicates that improving the partition heuristic alone is insufficient.

We then turn to runtime. In terms of runtime, \textsc{Rect3D} is
$2.69\times$ faster than FM+AR,
$6.55\times$ faster than FM+DAC23,
$3.26\times$ faster than Spectral+AR,
$6.79\times$ faster than Spectral+DAC23,
$15.98\times$ faster than GSP,
and $9.84\times$ faster than Lin.
This speedup primarily comes from keeping the optimization in continuous space
and using an efficient L-BFGS-B solver with a lightweight legalization step,
thereby avoiding expensive combinatorial moves and repeated repartitioning.

On the largest instance (n300), FM+DAC23 and Spectral+DAC23 do not finish within
the 1800\,s time limit, while the other four baselines yield much longer
wirelength and runtime; \textsc{Rect3D} completes in 42.1\,s with
$4.05\times 10^5$\,\si{\micro\meter} WL. Among the baselines that do finish,
the strongest competitor in WL is Spectral+AR, but it still requires
537.6\,s and produces $5.06\times 10^5$\,\si{\micro\meter} WL. In other words,
\textsc{Rect3D} is not merely trading runtime for better quality: on the
hardest benchmark, it simultaneously achieves better WL and substantially
shorter runtime than the best completed rectangular baseline.

Averages in the last row of~\Cref{tab:comparison_methods_doubled_runT} are computed
over all instances for which each method returns a solution (n10--n300 for FM+AR,
Spectral+AR, GSP, and Lin; n10--n200 for FM+DAC23 and Spectral+DAC23).
As the only method reported with five-run statistics, \textsc{Rect3D} additionally provides the mean and standard deviation to reflect solution stability.

Taken together, these results suggest that the advantage of \textsc{Rect3D} is
not tied to one specific benchmark or one specific baseline family. The method
outperforms both partition-first and combinatorial flows because it keeps die
assignment and geometric refinement coupled, while still solving the problem in
a smooth analytical space. This combination is precisely what allows
\textsc{Rect3D} to improve quality without paying the runtime cost usually
associated with richer 3D search spaces.

\subsection{Comparison with 3D Rectilinear Floorplanning Baselines}
\label{subsec:cmp_with_rectilinear}

To further evaluate \textsc{Rect3D} against 3D rectilinear floorplanning
baselines, we combine two partitioners, FM~\cite{2022TCAD-snap3d} and
spectral partitioning~\cite{1995DAC-Spectral_partitioning}, with three recent
2D rectilinear floorplanners, ISEDA~\cite{2025ISEDA_Rectilinear},
Jigsaw~\cite{2024ICCAD_Jisaw}, and Modern~\cite{2024ICCAD_Chen_Rectilinear_soft},
yielding six additional baselines: (1)~FM+ISEDA, (2)~FM+Jigsaw,
(3)~FM+Modern, (4)~Spectral+ISEDA, (5)~Spectral+Jigsaw, and
(6)~Spectral+Modern.

\noindent To ensure a fair comparison that isolates the effect of global
optimization quality, all six baselines share the same Rect3D legalization
(\Cref{alg:rect_legalization}). The only varying component is the 2D global
optimization kernel applied to each die after partitioning. This design removes
legalization as a confounding factor, so the remaining gap can be attributed to
the quality of the upstream optimization flow rather than to different
post-processing heuristics.

We first examine wirelength. \Cref{tab:comparison_rectilinear} shows that \textsc{Rect3D} achieves the best wirelength on every instance, by a wide margin.
Among the baselines, Spectral-based variants consistently outperform their FM counterparts, confirming that a better partition can noticeably improve the final layout quality.
The best rectilinear baseline overall is Spectral+ISEDA, which achieves an average WL ratio of 2.40$\times$ relative to \textsc{Rect3D}.
Spectral+Jigsaw is nearly identical (2.41$\times$), which is expected because both use the same ePlace-style engine and receive the same legalization.
FM+Modern performs worst (4.11$\times$), suggesting that its Adam-based optimizer with piecewise HPWL gradients is less effective than the ePlace density spreading used by ISEDA and Jigsaw. More importantly, even the best rectilinear baseline under the best partition remains far behind \textsc{Rect3D}, which shows that the dominant limitation is the partition-first decomposition itself rather than the particular 2D rectilinear engine.

We then turn to runtime. All partition-first baselines are faster than \textsc{Rect3D} on individual instances, with time ratios below 1 in \Cref{tab:comparison_rectilinear}. This is expected: each baseline optimizes a single die at a time after a fixed partition, whereas \textsc{Rect3D} jointly optimizes die assignment and geometry. However, this runtime advantage is modest relative to the quality gap. On n300, the strongest rectilinear baseline in WL, Spectral+ISEDA, completes in 32.45\,s but still produces $7.90\times 10^5$\,\si{\micro\meter} WL, nearly $1.95\times$ that of \textsc{Rect3D}. FM-based variants are faster still, but their average WL is about $3.9\times$--$4.1\times$ worse. The additional cost of joint 3D optimization is therefore well justified by the large improvement in final solution quality.

More importantly, both \Cref{tab:comparison_methods_doubled_runT} and \Cref{tab:comparison_rectilinear} show that fixing the die partition before layout optimization creates a strong ceiling on solution quality that 2D optimization alone can rarely overcome. Even pairing a state-of-the-art rectilinear optimizer with a good spectral partition still yields $2.4\times$ worse wirelength than \textsc{Rect3D}'s unified flow. This gap is especially pronounced on the larger instances (n200, n300), where cross-die interactions dominate the objective and a partition-oblivious approach often cannot recover.
This observation directly supports the motivation in \Cref{sec:intro}: extending 2D rectilinear floorplanning to 3D is not simply a matter of combining partitioning with a strong per-die rectilinear optimizer. Instead, the 3D problem requires native coordination between die assignment and in-die geometry, which is precisely what \textsc{Rect3D} preserves through its unified optimization flow.

\subsection{Effectiveness of Graph Laplacian Initialization}
\label{subsec:exp_of_init}
We next isolate the effect of the initialization stage. The graph Laplacian
initialization consistently leads to better final wirelength than the
grid-based alternative, showing that a better starting point still matters
after the full global optimization. To assess this effect, we conduct a
controlled comparison against a representative grid-based
initialization baseline, conceptually similar to the grid-oriented
rectilinear workflows used in prior work such as
Modern~\cite{2024ICCAD_Chen_Rectilinear_soft}. Both initialization
strategies are embedded into the same \textsc{Rect3D} framework,
ensuring that the global floorplanning procedure, regularization
schedule, and termination criteria are identical. Each result is
averaged over five runs.

As shown in~\Cref{tab:laplacian_vs_grid}, graph Laplacian initialization
consistently outperforms the grid-based baseline across all benchmarks. The
Laplacian method achieves an average HPWL ratio of 0.793 (20.7\%
improvement), with individual improvements ranging from 8.8\% (n50) to 31.2\%
(n100). On the larger benchmarks, such as n300, it delivers an absolute
improvement of 98,153\,\si{\micro\meter} while maintaining comparable
computational overhead (average time ratio 0.922).

The advantage of Laplacian initialization stems from its ability to encode
netlist connectivity directly into the initial embedding. Unlike grid-based
schemes that distribute blocks uniformly regardless of connectivity, the
Laplacian eigenvectors place strongly connected blocks close together from the
outset, giving the gradient-based optimizer a more structured landscape.

\begin{table}[t]
    \centering
    \tabcolsep=4pt
    \def\arraystretch{1.2}
    \footnotesize
    \caption{Comparison of representative grid-based and graph Laplacian initialization within \textsc{Rect3D}.}
    \label{tab:laplacian_vs_grid}
    \begin{threeparttable}
    \resizebox{\linewidth}{!}{%
    \begin{tabular}{c|cc|cc}
    \toprule
    \multirow{2}{*}{\textbf{Case}}
      & \multicolumn{2}{c|}{\textbf{Grid-based} (Modern-inspired~\cite{2024ICCAD_Chen_Rectilinear_soft})}
      & \multicolumn{2}{c}{\textbf{Laplacian} (ours)} \\
    \cmidrule(lr){2-3}
    \cmidrule(lr){4-5}
      & \textbf{Time} & \textbf{WL}
      & \textbf{Time} & \textbf{WL} \\
    \midrule
    n10  & 1.6$\pm$0.1 & 15521.2$\pm$4.4     & 1.1$\pm$0.0 & 11868.6$\pm$0.4 \\
    n30  & 14.0$\pm$0.7 & 39593.8$\pm$166.7   & 14.1$\pm$2.3 & 31214.7$\pm$232.3 \\
    n50  & 20.6$\pm$0.3 & 79172.7$\pm$568.1   & 18.1$\pm$1.3 & 72171.9$\pm$1171.4 \\
    n100 & 21.2$\pm$0.2 & 145121.9$\pm$347.4  & 21.0$\pm$0.4 & 99838.8$\pm$1048.8 \\
    n200 & 21.9$\pm$0.2 & 300536.5$\pm$2923.6 & 21.8$\pm$0.2 & 243236.8$\pm$25067.6 \\
    n300 & 22.9$\pm$0.2 & 483368.7$\pm$1387.9 & 22.8$\pm$0.2 & 385215.4$\pm$27404.1 \\
    \midrule
    Avg. ratio & 1 & 1 & \textbf{0.922} & \textbf{0.793} \\
    \bottomrule
    \end{tabular}
    }
    \begin{tablenotes}[flushleft]
    \footnotesize
    \item[*] Time is measured in seconds (s), WL in micrometers ($\mu$m). Values are\\mean $\pm$ standard deviation and each result is averaged over 5 runs.
    \end{tablenotes}
    \end{threeparttable}
\end{table}

This comparison also clarifies the role of initialization in the full
\textsc{Rect3D} pipeline. The later global optimization is powerful, but it is
not indifferent to where it starts. A topology-aware seed gives the optimizer a
more meaningful search direction from the beginning, which explains why the
benefit of the Laplacian stage remains visible even after all later phases are
allowed to converge.

\subsection{Case Study: Optimization Dynamics}
\label{subsec:flow_analysis}

\begin{figure*}[t]
    \centering
    \subfloat[Phase 1: Free Exploration]{%
        \raisebox{2pt}{\includegraphics[height=.17\linewidth]{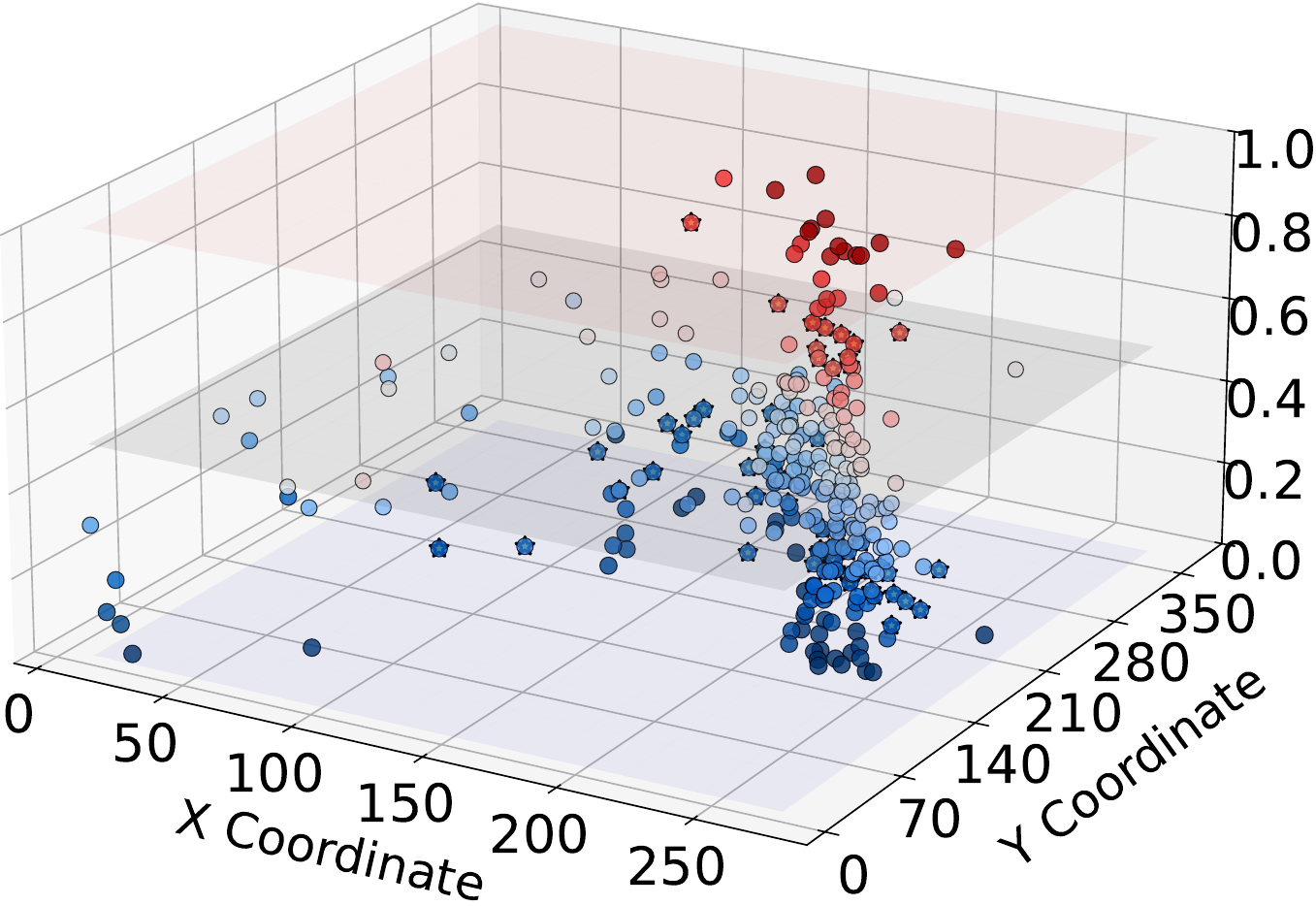}}%
        \label{fig:n300_procedure_phase1}
    }
    \subfloat[Phase 2: Layer Formation]{%
        \raisebox{2pt}{\includegraphics[height=.17\linewidth]{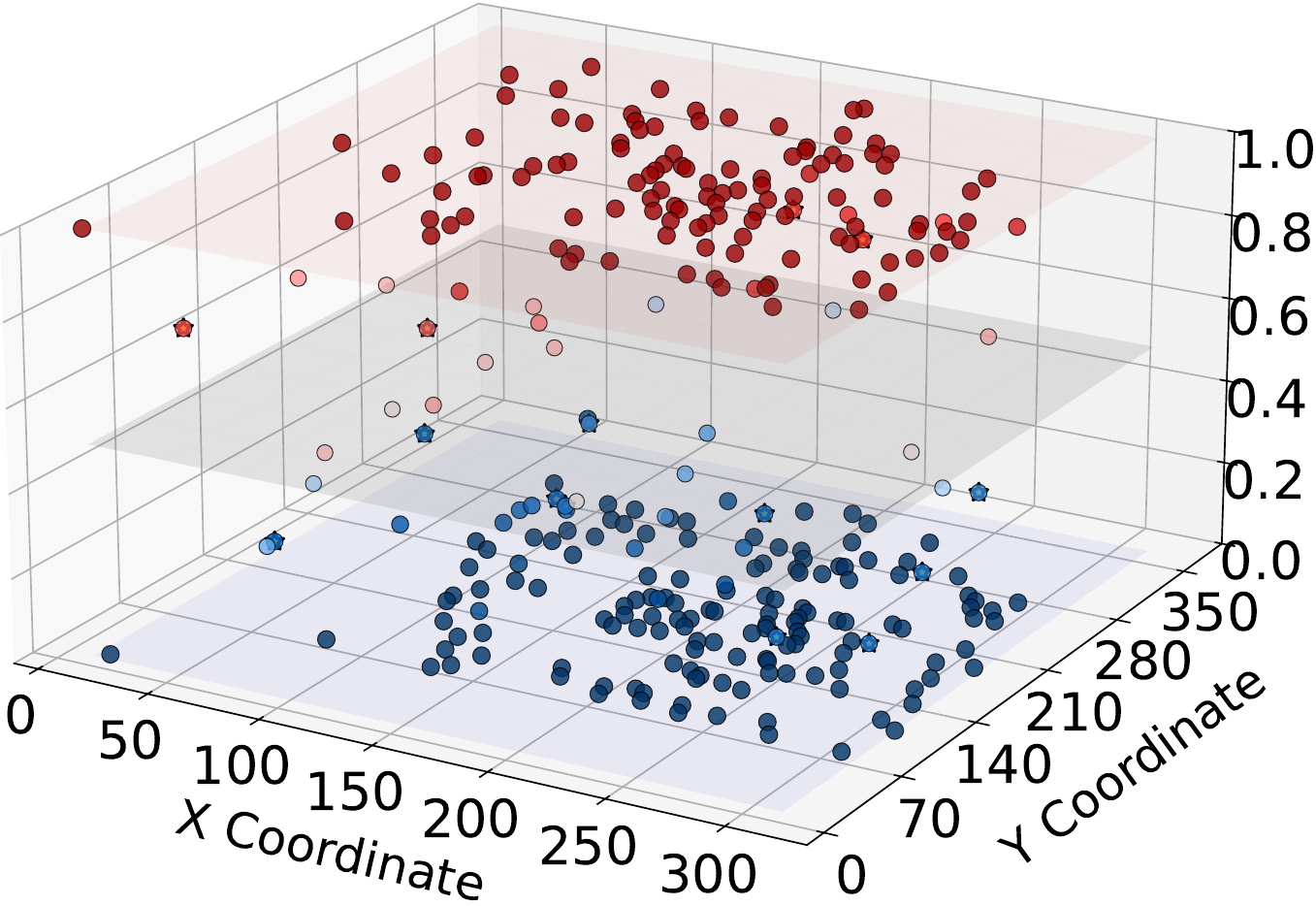}}%
        \label{fig:n300_procedure_phase2}
    }
    \subfloat[Phase 3: Die Convergence]{%
        \raisebox{2pt}{\includegraphics[height=.17\linewidth]{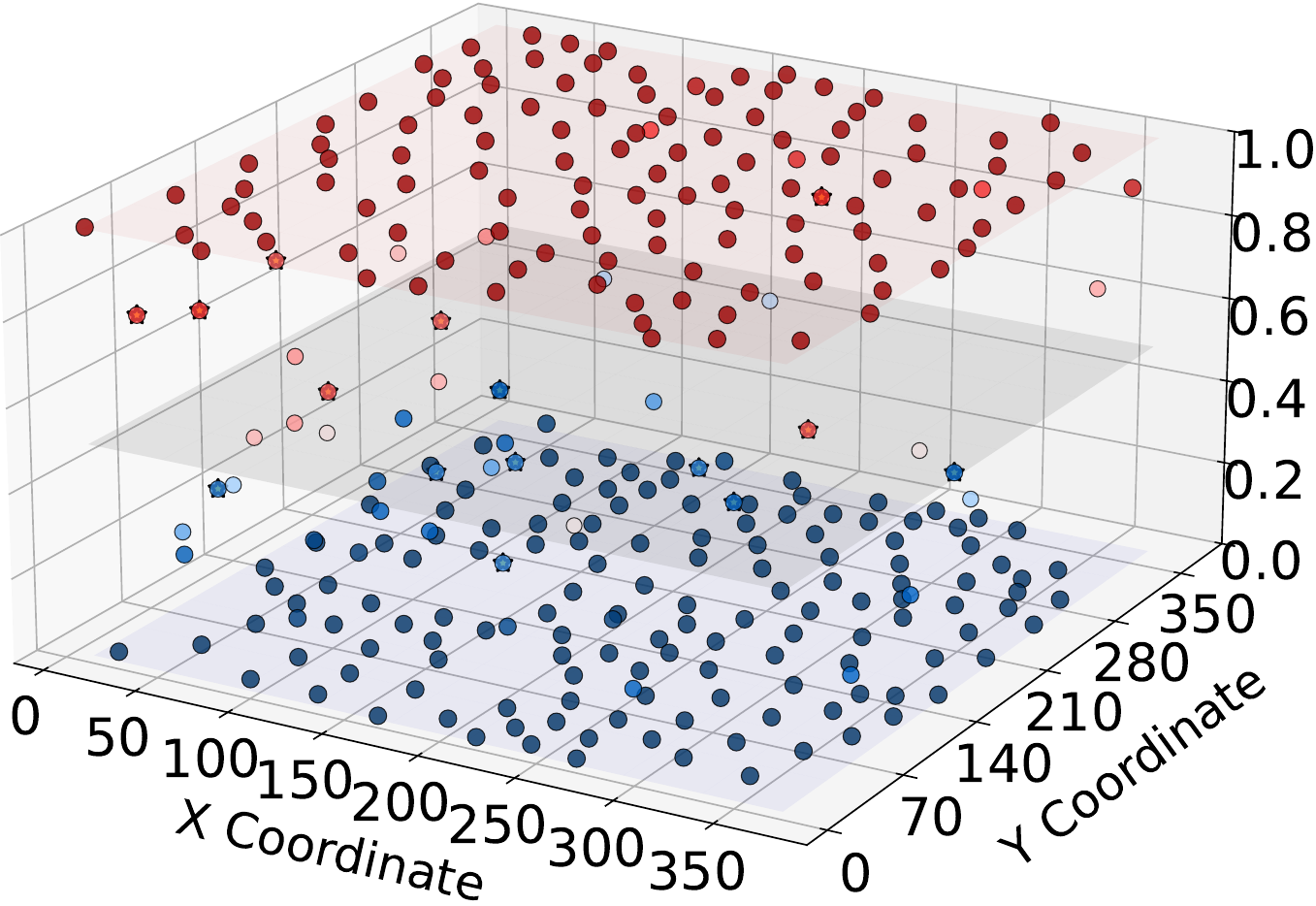}}%
        \label{fig:n300_procedure_phase3}
    }
    \subfloat[Convergence History]{%
        \includegraphics[height=.17\linewidth]{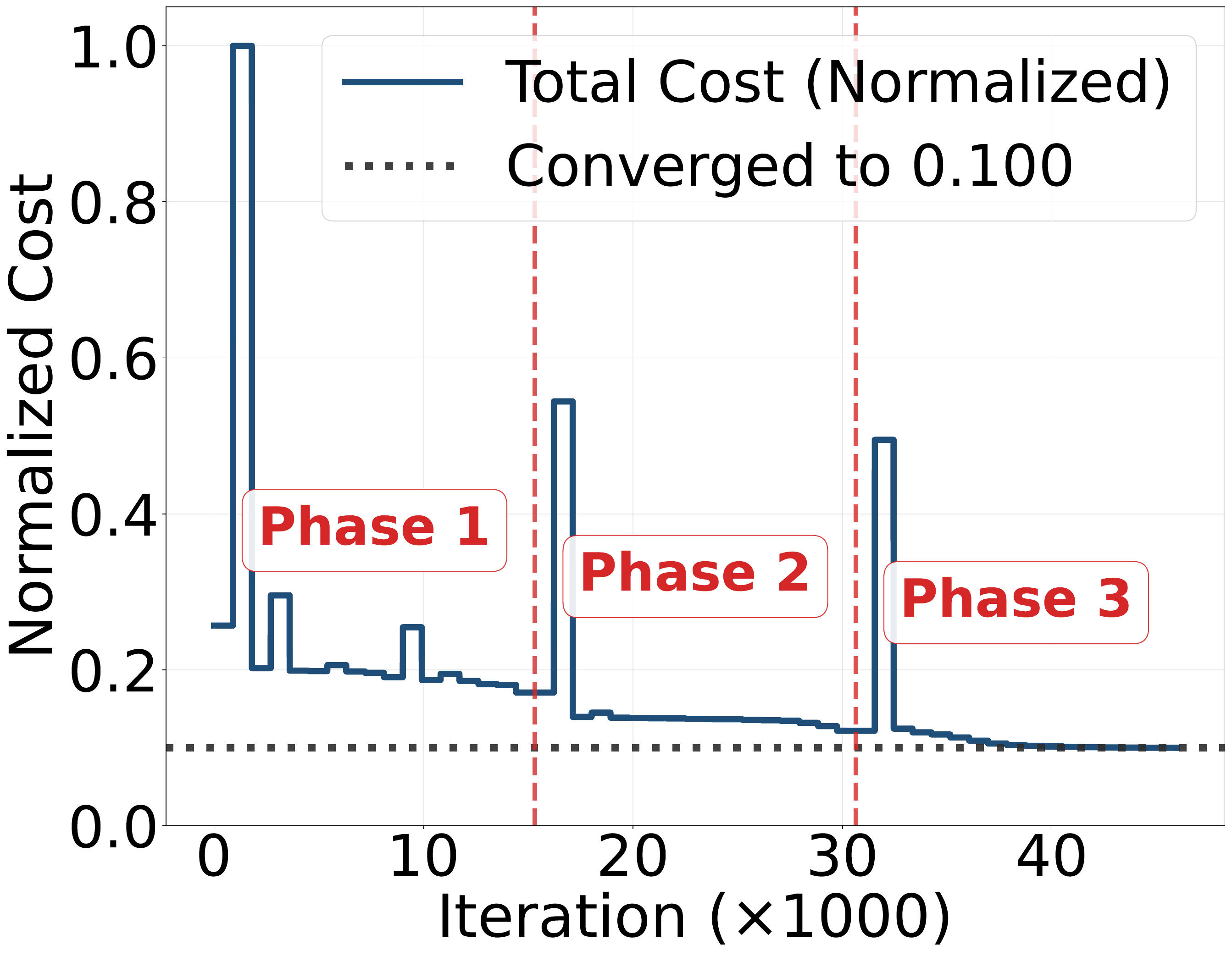}%
        \label{fig:n300_procedure_convergence_history}
    }
    \caption{\textsc{Rect3D} optimization procedure on the GSRC n300 benchmark. (a)--(c) show the progressive evolution of block distributions across the three optimization phases; each point represents a block, and color indicates the probabilistic die assignment (red: top die, blue: bottom die). (d) presents the corresponding convergence history of the normalized objective.}
    \label{fig:globalflp_flow}
\end{figure*}
\begin{figure*}[t]
    \centering
    \subfloat[n100 Top Die]{%
        \includegraphics[height=.145\linewidth]{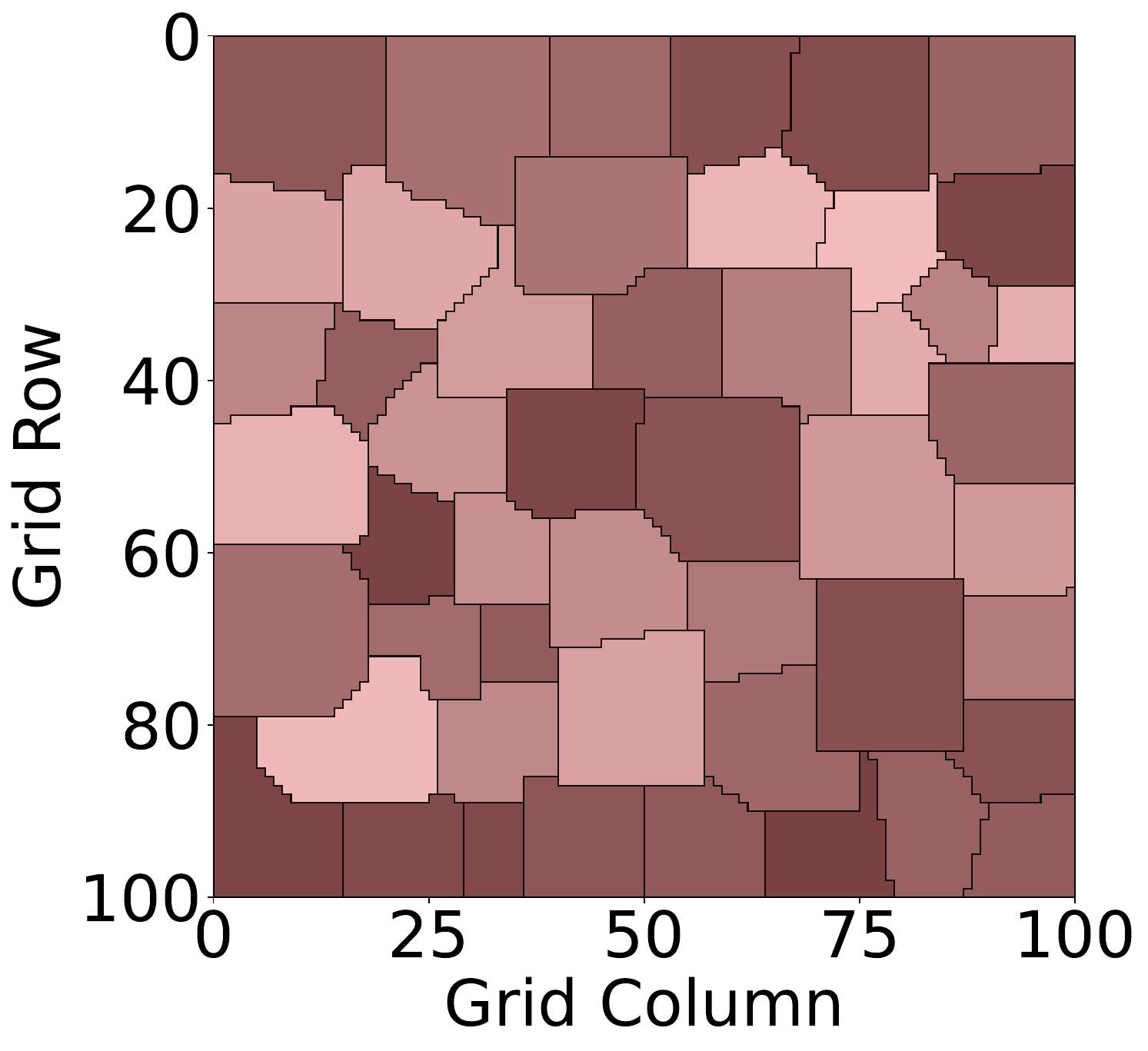}%
        \label{fig:n100_top}
    }
    \hfill
    \subfloat[n100 Bottom Die]{%
        \includegraphics[height=.145\linewidth]{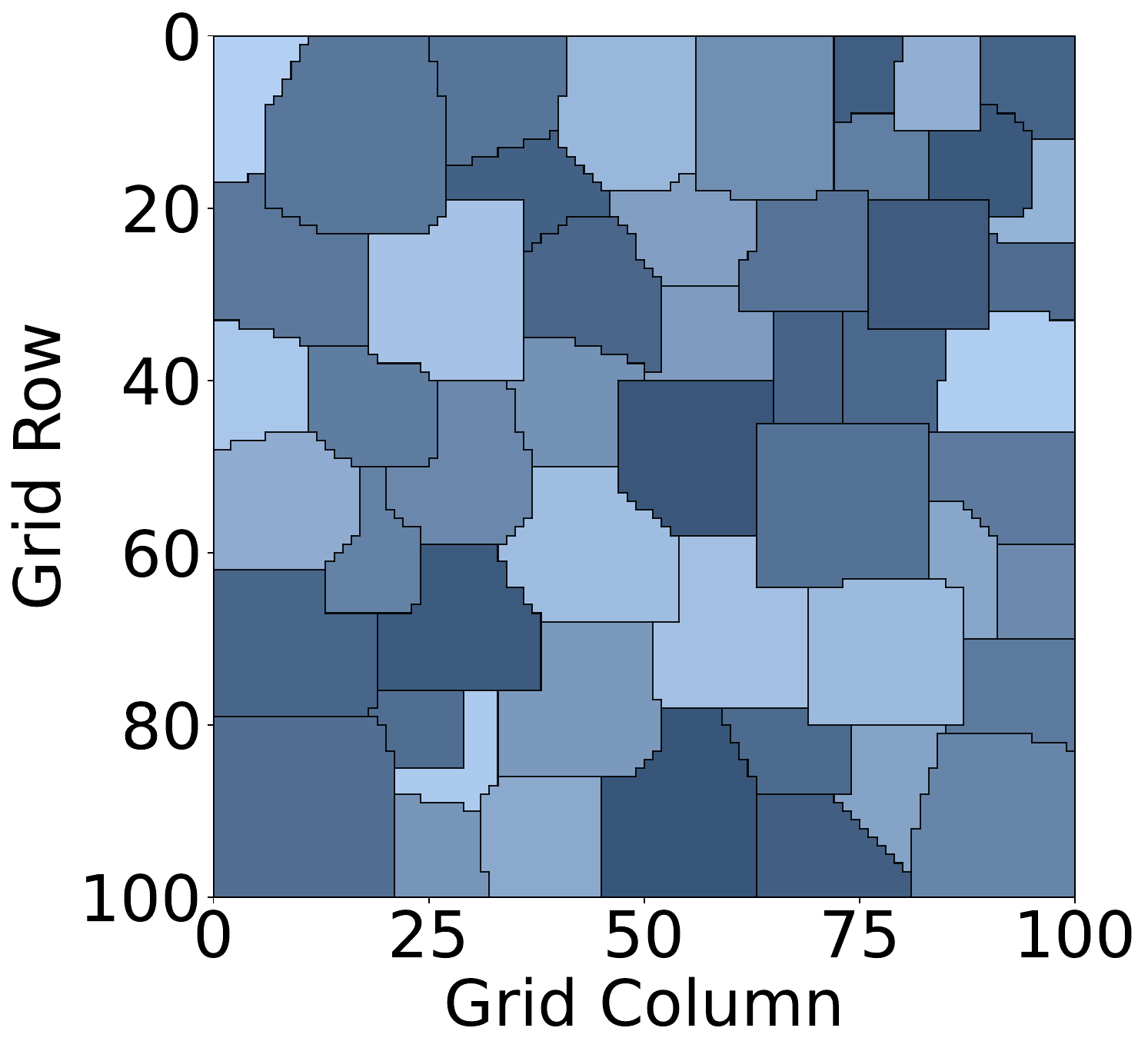}%
        \label{fig:n100_bottom}
    }
    \hfill
    \subfloat[n200 Top Die]{%
        \includegraphics[height=.145\linewidth]{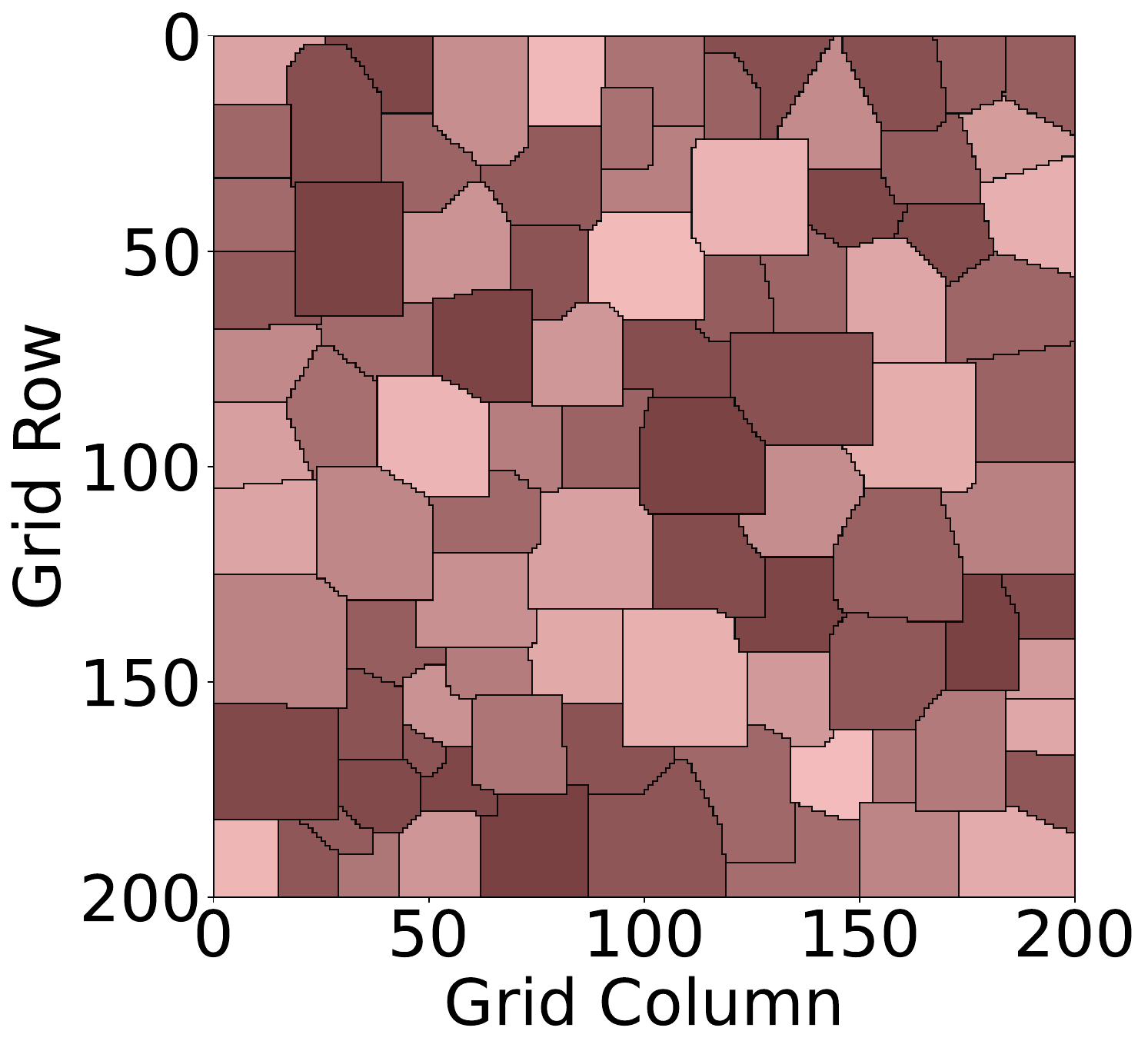}%
        \label{fig:n200_top}
    }
    \hfill
    \subfloat[n200 Bottom Die]{%
        \includegraphics[height=.145\linewidth]{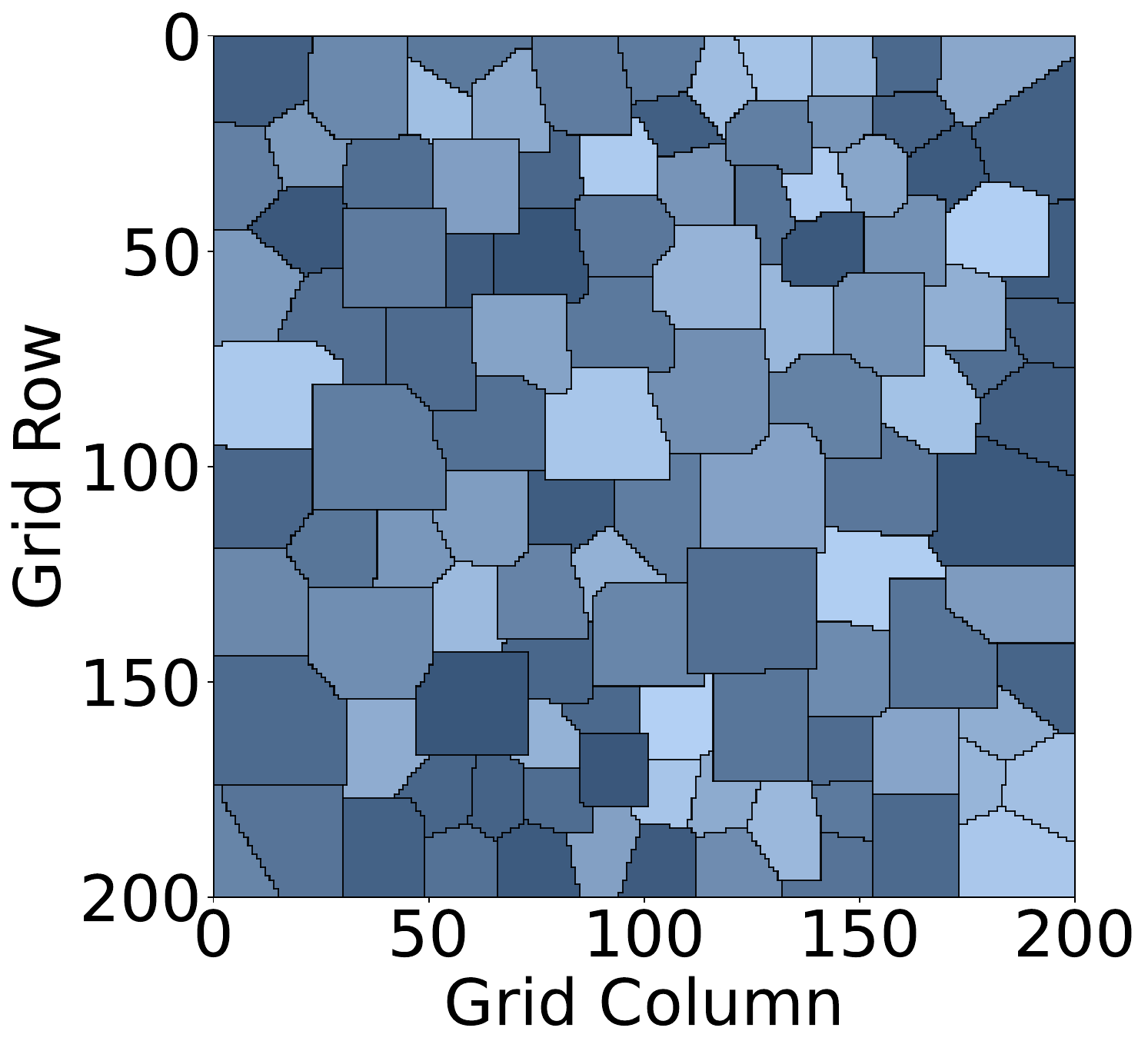}%
        \label{fig:n200_bottom}
    }
    \hfill
    \subfloat[n300 Top Die]{%
        \includegraphics[height=.145\linewidth]{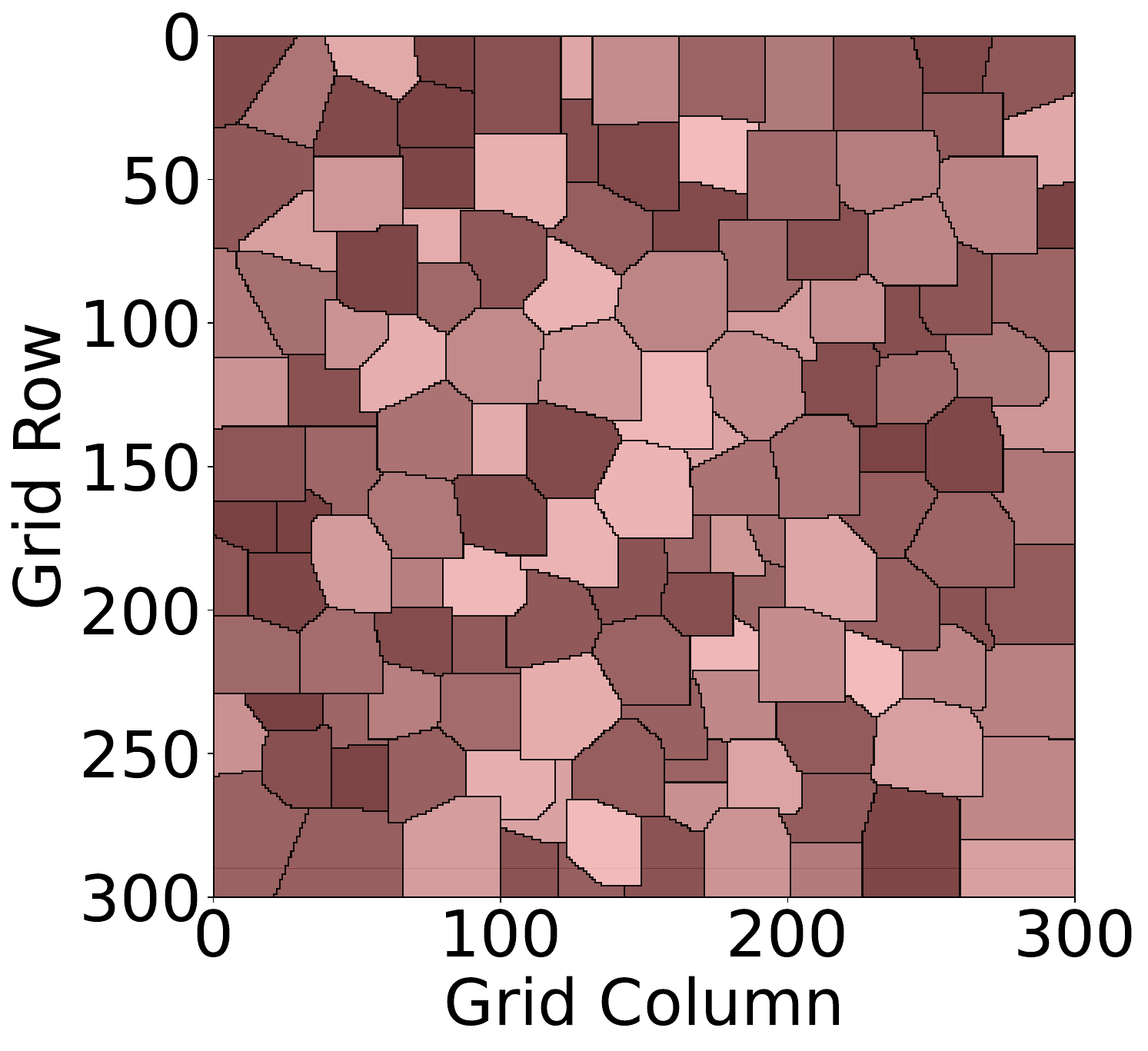}%
        \label{fig:n300_top}
    }
    \hfill
    \subfloat[n300 Bottom Die]{%
        \includegraphics[height=.145\linewidth]{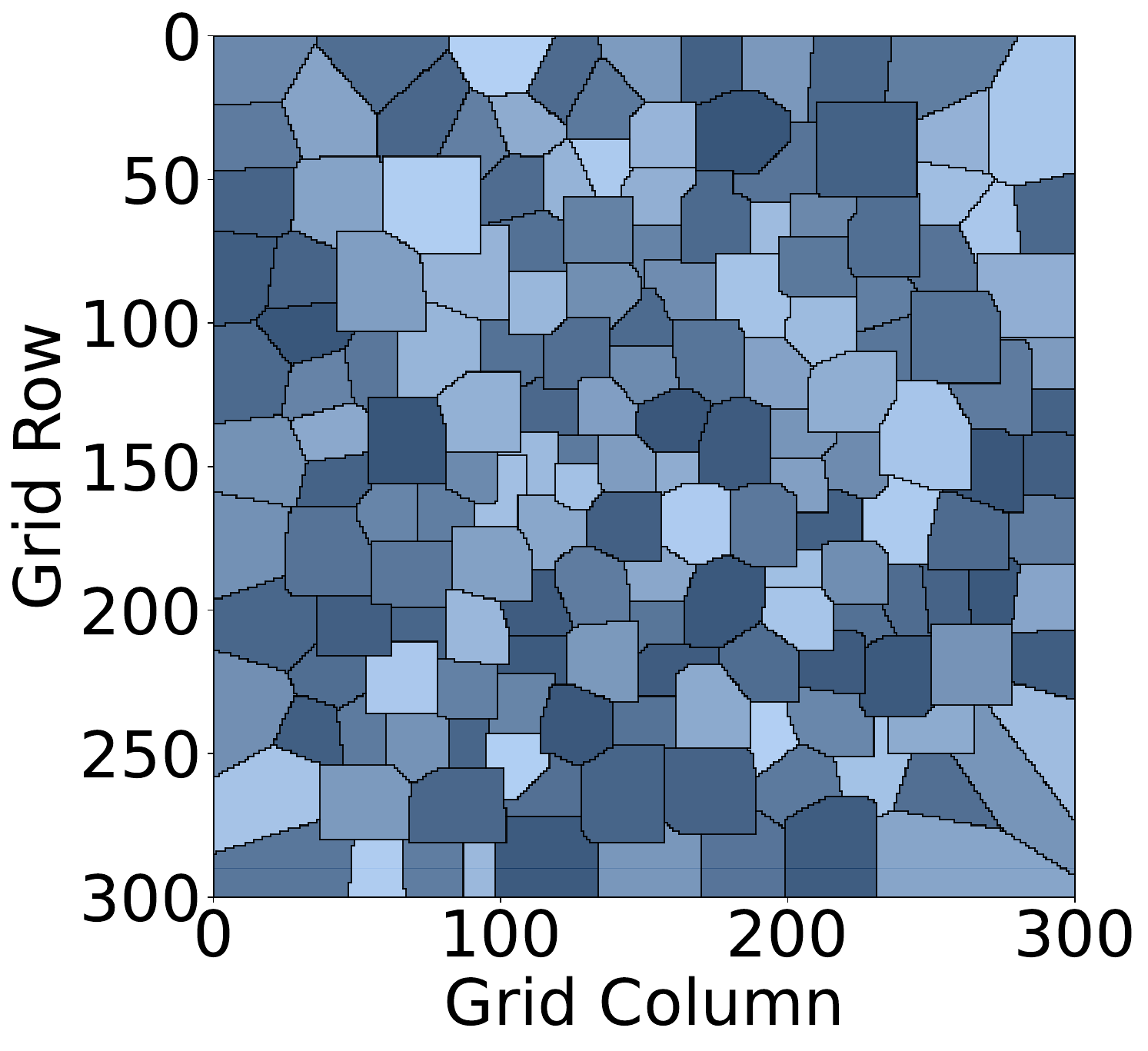}%
        \label{fig:n300_bottom}
    }
    \caption{3D rectilinear floorplans produced by \textsc{Rect3D} on GSRC benchmarks n100, n200, and n300. Each colored region represents a block's rectilinear footprint on the assigned die. The layouts demonstrate dense, overlap-free packing with contiguous rectilinear shapes across different problem scales.}
    \label{fig:gsrc_3dflp_all}
\end{figure*}

\begin{figure}[t!]
    \centering
    \includegraphics[width=.95\linewidth]{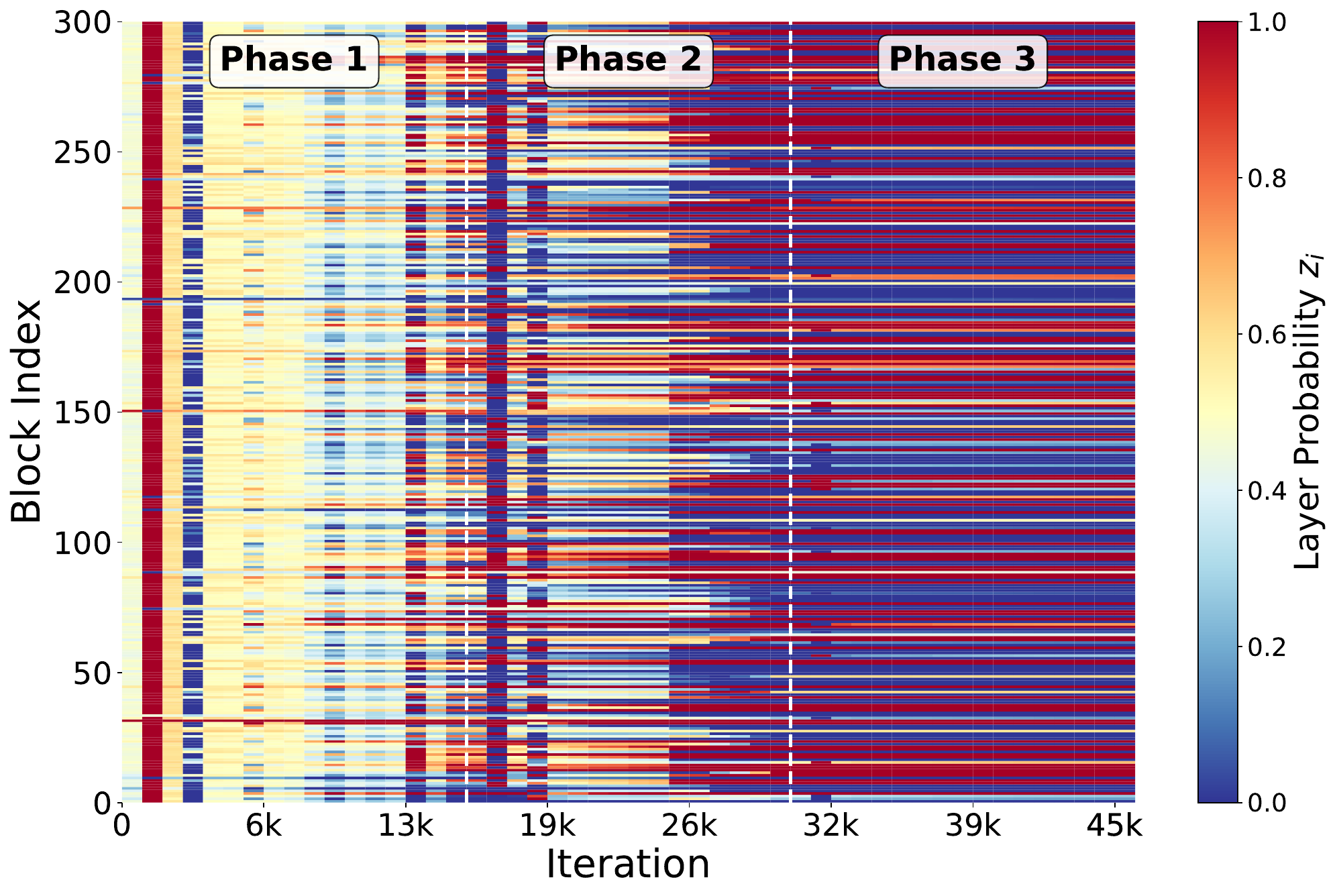}
    \caption{Per-block $z$-coordinate trajectories on n300 (rows: blocks, columns: iterations). Dashed lines mark phase boundaries. Notably, blocks do not commit monotonically to one die: many reverse their tendency across phases, reflecting the optimizer's flexible exploration of the solution space. These reversals coincide with the objective spikes observed at phase transitions in~\Cref{fig:n300_procedure_convergence_history}.}
    \label{fig:z_evolution_heatmap}
\end{figure}

We next examine the GSRC n300 benchmark to show how the three-phase schedule
behaves during optimization. The case study reveals a clear progression from
free exploration to layer formation and finally to die convergence. This
behavior is also consistent with the Stage-2 design shown in the middle panel
of~\Cref{fig:framework}: expected overlap is initially high when many blocks
have ambiguous $z$-values, then decreases as two layers emerge, and finally
reaches a fully separated top/bottom assignment before legalization.

\Cref{fig:globalflp_flow} visualizes the \textsc{Rect3D} optimization procedure.
Each circular point denotes the probabilistic location of a block during
optimization, where red and blue correspond to the likelihood of being assigned
to the top or bottom die.

In Phase~1 (\Cref{fig:n300_procedure_phase1}), the regularization is weak
($\alpha = \alpha_{\min}$), so blocks explore the 3D space with few layer
restrictions. Many blocks appear in intermediate colors, which indicates
ambiguous die assignments with $z_i \approx 0.5$. At this point, the optimizer
mainly reduces wirelength and roughly satisfies the overlap constraints, thus
building a coarse spatial layout.

In Phase~2 (\Cref{fig:n300_procedure_phase2}), the regularization weight is
increased to $\alpha_{\text{mid}}$. The penalty term~\Cref{eq:z_regularization}
then pushes the $z$-values toward 0 or 1, and two clusters begin to form for
the top and bottom dies. Blocks with strong same-die connections usually settle
earlier, while blocks with more cross-die interactions take longer.

In Phase~3 (\Cref{fig:n300_procedure_phase3}), the regularization reaches
$\alpha_{\max}$ and nearly all blocks converge to binary die assignments,
showing pure red or blue colors. The two dies are now clearly separated, and
each die contains a dense block arrangement.

The per-block trajectories in~\Cref{fig:z_evolution_heatmap} further confirm
this interpretation. During Phase~1, many rows remain near the middle range,
indicating deliberately ambiguous die assignments while the optimizer searches
for favorable planar structure. In Phase~2, the trajectories begin to polarize
into two bands as layer formation becomes stronger. In Phase~3, most rows
stabilize near 0 or 1, which matches the final hard die separation illustrated
in the Stage-2 part of~\Cref{fig:framework}. The fact that some trajectories
reverse direction across phase boundaries is also consistent with the framework
design: the method is intended to postpone irreversible die decisions until
enough geometric context has been established.

The convergence curve in~\Cref{fig:n300_procedure_convergence_history} plots
the normalized unified objective~\Cref{eq:unified_3d_objective}. The sharp rise
at the beginning of Phase~1 comes from the initialization $\mathbf{X}_0$
defined by~\Cref{eq:3d_coordinate_init,eq:coordinate_standardization}. As
explained in~\Cref{subsec:graph_laplacian_initialization}, the Laplacian
embedding relaxes non-overlap to orthonormality, so the initial solution
incurs large constraint penalties under the full objective. Similar spikes
appear at later phase transitions, when the stronger regularization changes the
objective before the optimizer adapts. These oscillations are expected rather
than problematic. Each phase change increases the weight of die-assignment
regularization and therefore changes the effective objective landscape. A
layout that is favorable under weak regularization may no longer remain
favorable once the optimizer is asked to produce a clearer top/bottom
separation. As a result, the objective can rise or fluctuate briefly while the
block coordinates and die variables adjust to the new balance among wirelength,
overlap, outline, and regularization terms. After this short adjustment, the
objective decreases again and converges to the value marked by the dashed line.
This behavior shows that the three-phase schedule performs controlled
re-optimization rather than unstable search.

\subsection{Legalization Results}
\label{subsec:legalization_results}

After global optimization, the continuous solution must still be converted into
a legal rectilinear floorplan. The results in~\Cref{fig:gsrc_3dflp_all} show
that the legalization stage preserves the global structure while producing
feasible layouts for benchmarks n100, n200, and n300.

Several observations follow from the legalized layouts. All blocks form
contiguous rectilinear regions without holes or disconnected fragments, which
confirms the effectiveness of the connectivity-preserving CA refinement in
~\Cref{alg:rect_legalization}. The die area is also well utilized on both
dies, with little whitespace. The layouts scale well: even the n300 benchmark
with 300 blocks produces a clean and non-overlapping floorplan on both dies
within 42.1\,s total runtime.

These final 3D rectilinear floorplans satisfy all constraints in~\Cref{eq:3d_formulation}: non-overlap (\Cref{eq:formulation_constraint_nonoverlap}), area and aspect ratio (\Cref{eq:formulation_constraint_shape}), and outline boundary (\Cref{eq:formulation_constraint_outline_boundary}).

\section{Conclusion}
\label{sec:conclusion}
In this paper, we presented \textsc{Rect3D}, a unified analytical framework for
3D-IC rectilinear floorplanning that avoids separating die assignment from
intra-die layout too early. \textsc{Rect3D} combines a topology-aware
initialization, a unified global optimization with probabilistic die
assignment, and a legalization stage that converts the result into connected
rectilinear layouts. This design preserves cross-die flexibility during
optimization while still producing legal floorplans efficiently.
Experimental results on the GSRC benchmark suite show that \textsc{Rect3D}
reduces wirelength by 46.8\%--83.6\% and improves runtime by
$2.69\times$--$15.98\times$ over six representative baselines, while remaining
effective on instances with up to 300 blocks. Together with its consistent
wirelength advantage over additional partition-first rectilinear baselines,
these results show that jointly optimizing die assignment and in-die geometry
is a promising direction for scalable 3D rectilinear floorplanning. Future
work includes extending the framework to thermal-aware objectives and
multi-tier stacking.

\bibliographystyle{IEEEtran}
\bibliography{ref/Top,ref/reference}

\vspace{-.2in}
\begin{IEEEbiography}
    [{\includegraphics[width=1.0in,height=1.26in,clip,keepaspectratio]{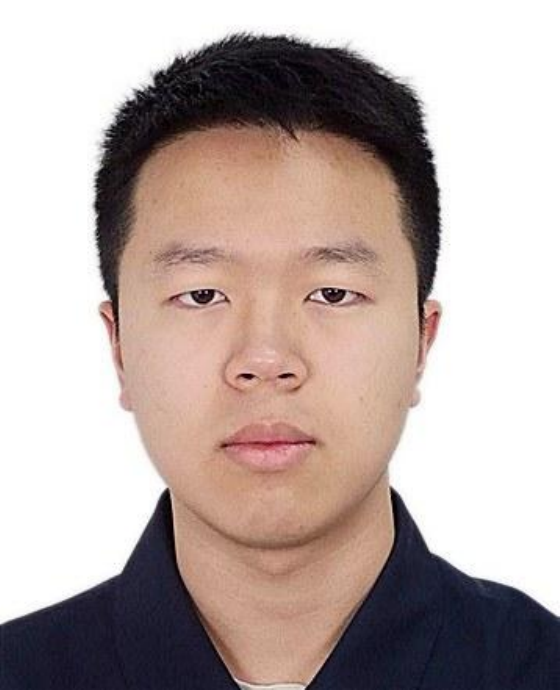}}]
    {Shuo~Ren}
    received his B.S. degree in computer science from the Huazhong University of Science and Technology, Wuhan, China, in 2024. He is currently working toward the Ph.D. degree in the Department of Computer Science and Engineering, The Chinese University of Hong Kong. His current research focuses on electronic design automation, with particular interests in 3D integrated circuit design and physical design optimization.
\end{IEEEbiography}

\vspace{-.2in}
\begin{IEEEbiography}
    [{\includegraphics[width=1.0in,height=1.26in,clip,keepaspectratio]{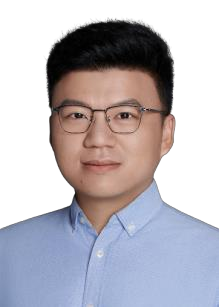}}]
    {Rongliang~Fu}
    received his Ph.D. in Computer Science and Engineering from The Chinese University of Hong Kong in January 2026, following an M.S. from the University of Chinese Academy of Sciences in June 2021 and a B.S. in Software Engineering from Northwestern Polytechnical University in June 2018. He has authored over 30 papers across major journals (IEEE TC and IEEE TCAD) and conferences (DAC, DATE, ICCAD, etc.). His research spans electronic design automation and EDA for superconducting electronics.
\end{IEEEbiography}

\vspace{-.2in}
\begin{IEEEbiography}
    [{\includegraphics[width=1.0in,height=1.26in,clip,keepaspectratio]{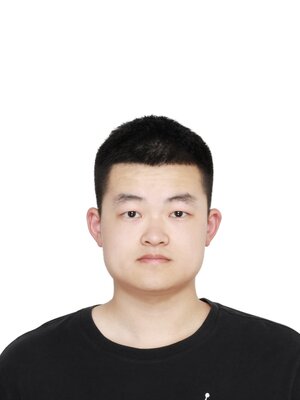}}]
    {Libo~Shen}
    received his B.S. degree in communication engineering from Beijing University of Posts and Telecommunications, Beijing, China, in 2021 and his M.S. degree in computer technology from the University of Chinese Academy of Sciences, Beijing, China, in 2024. He is currently a Ph.D. student at the Department of Computer Science and Engineering, The Chinese University of Hong Kong. His research interests include electronic design automation and computer architecture.
\end{IEEEbiography}

\vspace{-.2in}
\begin{IEEEbiography}
    [{\includegraphics[width=1.0in,height=1.26in,clip,keepaspectratio]{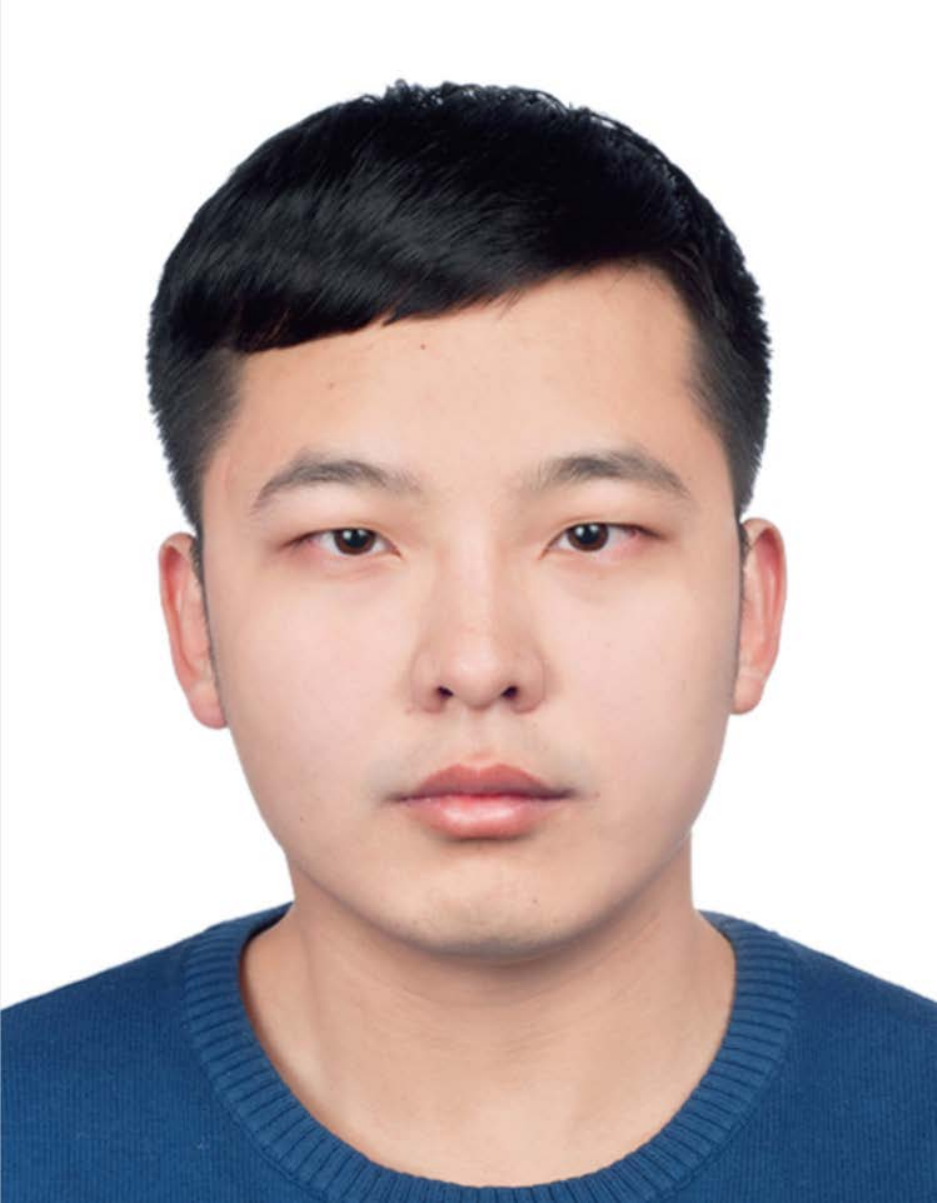}}]
    {Zhen~Zhuang}
    received his M.Eng. and B.Eng. from Fuzhou University in 2021 and 2018, respectively. He obtained his Ph.D. from the Chinese University of Hong Kong in 2025 under the supervision of Prof. Tsung-Yi Ho. His current research interest is Electronic Design Automation (EDA), especially EDA for advanced packaging and 3D IC. He was a recipient of three ICCAD/ISPD contest awards.
\end{IEEEbiography}

\vspace{-.2in}
\begin{IEEEbiography}
    [{\includegraphics[width=1.0in,height=1.26in,clip,keepaspectratio]{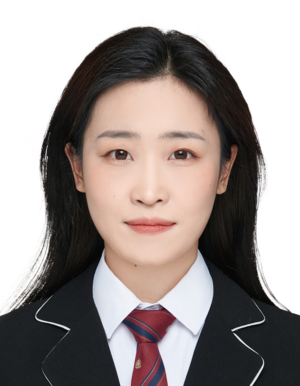}}]
    {Leilei~Jin}
    is currently a postdoctoral researcher affiliated with the Department of Computer Science and Engineering at The Chinese University of Hong Kong (CUHK). She received her Ph.D. degree from Southeast University in 2024, where her doctoral work focused on foundational theories and practical methodologies in integrated circuit design automation. Her research interests include static timing analysis, crosstalk prediction, PPA optimization for Backside PDN and 3DICs under advanced process nodes.
\end{IEEEbiography}

\vspace{-.2in}
\begin{IEEEbiography}
    [{\includegraphics[width=1.0in,height=1.26in,clip,keepaspectratio]{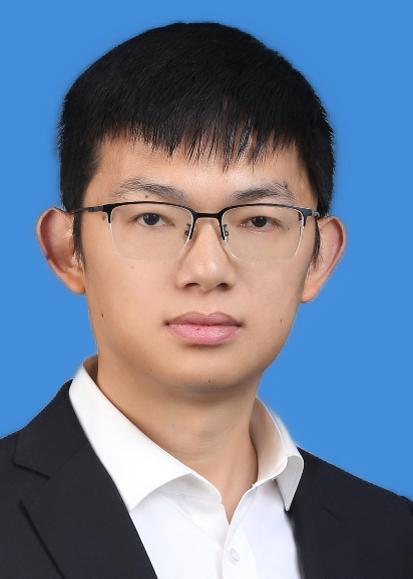}}]
    {Chen~Wu} received the B.S. degree from the University of Electronic Science and Technology of China, Chengdu, China, in 2012; the M.S. degree from Tsinghua University, Beijing, China, in 2015; and the Ph.D. degree from the University of California, Los Angeles (UCLA), CA, USA. He is now a research fellow at Ningbo Institute of Digital Twin, Eastern Institute of Technology, Ningbo, China.
    His current research interests include AI computer architecture, AI compiler,  Shift-Lefted EDA for Chiplet, and AI4EDA.
\end{IEEEbiography}

\vspace{-.2in}
\begin{IEEEbiography}
    [{\includegraphics[width=1.0in,height=1.26in,clip,keepaspectratio]{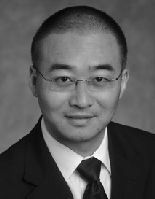}}]
    {Lei~He} (M'03-SM'08-F'24) received the Ph.D. degree in computer science from the University of California, Los Angeles (UCLA), CA, USA, in 1999. He was a Faculty Member with the University of Wisconsin, Madison, WI, USA, from 1999 to 2002. He consulted Cadence Design Systems, Cisco, Empyrean Soft, HewlettPackage, Intel, and Synopsys, and was a Founding Technical Advisory Board Member for Apache Design Solutions and Rio Design Automation. He is a Co-Founder of Silicon Cloud International and the Founder of NxEco, Inc. He is currently a Professor with the Electrical and Computer Engineering Department, UCLA; a Chair Professor with the Eastern Institute of Technology, Ningbo, China. He has authored or coauthored one book and more than 200 technical papers. His current research interests include modeling and simulation, very-large-scale integration circuits and systems, Internet-of-things, and artificial intelligence. Dr. He was a recipient of many best paper nominations and awards, including the 2010 ACM Transactions on Electronic System Design Automation Best Paper Award and the 2011 IEEE Circuit and System Society Darlington Best Paper Award.
\end{IEEEbiography}

\vspace{-.4in}
\begin{IEEEbiography}
    [{\includegraphics[width=1.0in,height=1.26in,clip,keepaspectratio]{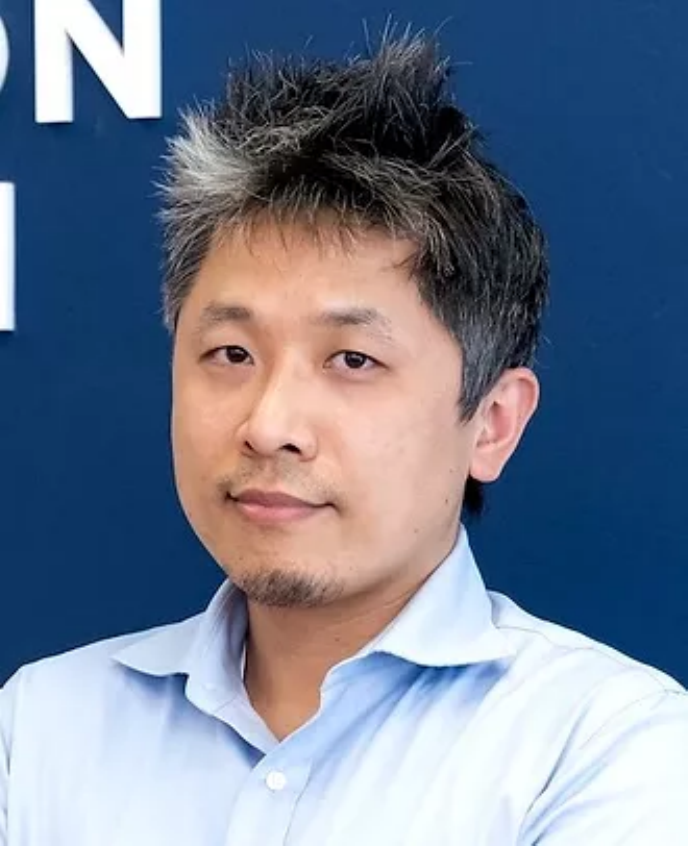}}]
    {Bei~Yu}
    (M'15-SM'22)
    received the Ph.D.~degree from The University of Texas at Austin in 2014.
    He is currently a Professor in the Department of Computer Science and Engineering, The Chinese University of Hong Kong.
    He has served as TPC Chair of ACM/IEEE Workshop on Machine Learning for CAD, and in many journal editorial boards and conference committees.
    He received ten Best Paper Awards from IEEE TSM 2022, DATE 2022, ICCAD 2021 \& 2013, ASPDAC 2021 \& 2012, ICTAI 2019, Integration, the VLSI Journal in 2018, ISPD 2017, SPIE Advanced Lithography Conference 2016, and many other awards, including DAC Under-40 Innovator Award (2024), IEEE CEDA Ernest S.~Kuh Early Career Award (2022), and Hong Kong RGC Research Fellowship Scheme (RFS) Award (2024).
\end{IEEEbiography}

\vspace{-.4in}
\begin{IEEEbiography}
    [{\includegraphics[width=1.0in,height=1.26in,clip,keepaspectratio]{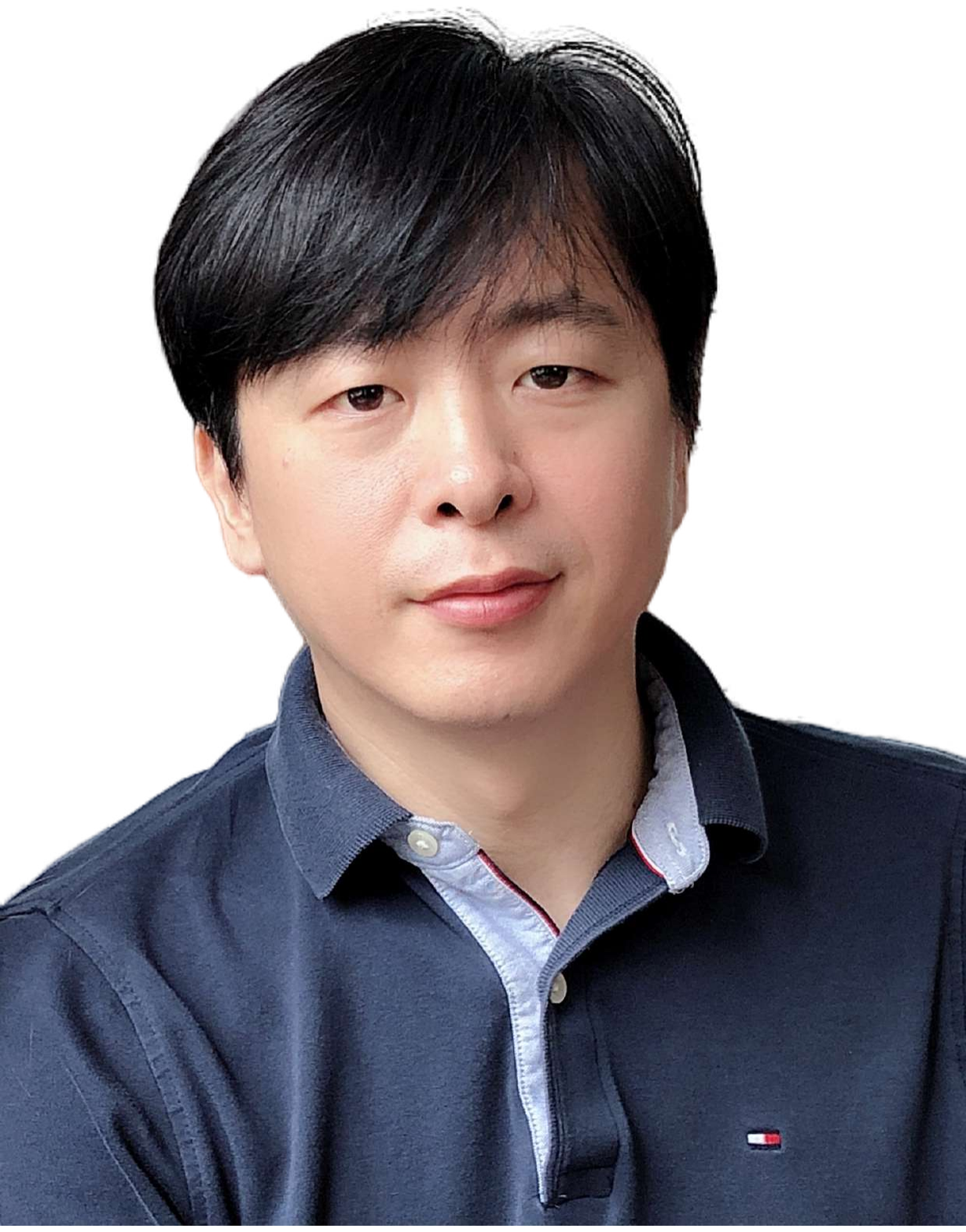}}]
    {Tsung-Yi~Ho}
    (F'24)
    is a Professor in the Department of Computer Science and Engineering, The Chinese University of Hong Kong (CUHK). He received his Ph.D. in Electrical Engineering from National Taiwan University in 2005. His research interests include several areas of computing and emerging technologies, especially in the design automation of microfluidic biochips. He was a recipient of the Best Paper Award at the IEEE Transactions on Computer-Aided Design of Integrated Circuits and Systems in 2015. Currently, he serves as the VP Conferences of IEEE CEDA, and the Executive Committee of ASP-DAC and ICCAD. He is a Distinguished Member of ACM and a Fellow of IEEE.
\end{IEEEbiography}

\end{document}